\documentclass[hidelinks, reprint, amsmath,amssymb, aps]{revtex4-2}

\usepackage{hyperref}
\usepackage{orcidlink}
\usepackage{graphicx}
\usepackage{dcolumn}
\usepackage{bm}
\usepackage{physics}
\usetikzlibrary{quantikz2}
\usepackage{amssymb}

\definecolor{pistacchio}{HTML}{C1D37F}
\definecolor{periwinkle}{HTML}{AFBBF2}
\definecolor{mutedplum}{HTML}{7B3F61}
\definecolor{iceblue}{HTML}{88D9E6}

\begin{document}

\title{Measurement-based quantum computing with qudit stabilizer states}

\author{Alena Romanova \orcidlink{0000-0003-0771-8835}}
 \email{corresponding author: alena.romanova@uibk.ac.at}
\author{Wolfgang Dür \orcidlink{0000-0002-0234-7425}}
 \email{wolfgang.duer@uibk.ac.at}
\affiliation{
 Universität Innsbruck, Institut für Theoretische Physik \\ Technikerstraße 21a, 6020 Innsbruck, Austria \\
}

\date{\today}

\begin{abstract}
We show how to perform measurement-based quantum computing on qudits (high-dimensional quantum systems) using alternative resource states beyond the cluster state. Estimating overheads for gate decomposition, we find that generalizing standard qubit measurement patterns to the qudit cluster state is suboptimal in most dimensions, so that alternative qudit resource states could enable enhanced computational efficiency. In these resources, the entangling interaction is a block-diagonal Clifford operation rather than the usual controlled-phase gate for cluster states. This simple change has remarkable consequences: the applied entangling operation determines an intrinsic single-qudit gate associated with the resource that drives the quantum computation when performing single-qudit measurements on the resource state. We prove a condition for the intrinsic gate allowing for the measurement-based implementation of arbitrary single-qudit unitaries. Furthermore, we demonstrate for prime-power-dimensional qudits that the complexity of the realization depends linearly both on the dimension and the Pauli order of the intrinsic gate. These insights also allow us to optimize the efficiency of the standard qudit cluster state by effectively mimicking more favorable intrinsic-gate structures, thereby reducing the required measurement depth. Finally, we discuss the required two-dimensional geometry of the resource state for universal measurement-based quantum computing. As concrete examples, we present multiple alternative resource states. In certain dimensions, we show the existence of resource states achieving optimal intrinsic gates, enabling more efficient measurement-based quantum information processing than the qudit cluster state and highlighting the potential of qudit stabilizer state resources for future quantum computing architectures.
\end{abstract}

\maketitle

\section{Introduction}

\begin{figure*}[t!]
    \centering
    \begin{tikzpicture}

    \node at (-4,0.5) {$(a)$};
    
    \coordinate (1) at (-2.25,0);  
    \coordinate (2) at (-0.75,0); 

    \draw[thick] (1) -- (2);

    \foreach \i in {1,2} {
        \node[circle, fill=pistacchio, draw=black] at (\i) {};
    }
    \node at (-2.25,0.5) {$\ket{\psi}$};

    \node at (-4,-1) {$(b)$};
    
    \coordinate (3) at (-3,-1.5);  
    \coordinate (4) at (-1.5,-1.5); 
    \coordinate (5) at (0,-1.5);

    \fill[periwinkle, opacity=0.3, rounded corners=5pt] 
    (-3.4,-1.85) rectangle (0.4,-1.15);
    
    \draw[thick] (3) -- (5);

    \foreach \i in {3,4,5} {
        \node[circle, fill=pistacchio, draw=black] at (\i) {};
    }

    \node at (-4,-2.5) {$(c)$};
    
    \coordinate (7) at (-3,-3); 
    \coordinate (8) at (-1.5,-3); 
    \coordinate (9) at (0,-3); 

    \coordinate (10) at (-3,-4); 
    \coordinate (11) at (-1.5,-4); 
    \coordinate (12) at (0,-4);

    \fill[iceblue, opacity=0.3, rounded corners=5pt]
    (-1.9,-4.4) rectangle (-1.1,-2.6);
    
    \draw[thick] (7) -- (9);
    \draw[thick] (10) -- (12);
    \draw[thick] (8) -- (11);

    \foreach \i in {7,8,9,10,11,12} {
        \node[circle, fill=pistacchio, draw=black] at (\i) {};
    }

    \draw [decorate,decoration={brace,amplitude=5pt,mirror,raise=10pt}]
    (7) -- (10) node [black,midway,xshift=-0.8cm] 
    {$\ket{\psi}$};

    \node at (1.5,0.5) {$(d)$};
        
    \foreach \x in {2.5, 4, 5.5,7,8.5,10,11.5} {
        \foreach \y in {0,-2,-3,-4} {
            \coordinate (\x\y) at (\x,\y);
        }
    }

    \fill[periwinkle, opacity=0.3, rounded corners=5pt] 
    (2,-0.5) rectangle (3,0.5);
    \fill[periwinkle, opacity=0.3, rounded corners=5pt] 
    (2,-2.5) rectangle (3,-1.5);
    \fill[iceblue, opacity=0.3,rounded corners=5pt] 
    (3.5,-2.5) rectangle (4.5, 0.5);

    \fill[periwinkle, opacity=0.3, rounded corners=5pt] 
    (5,-2.5) rectangle (7.5,-1.5);
    \fill[periwinkle, opacity=0.3, rounded corners=5pt] 
    (9.5,-1.5) rectangle (10.5,-2.5);
    \fill[iceblue, opacity=0.3, rounded corners=5pt] 
    (8,-4.5) rectangle (9,-1.5);

    \fill[periwinkle, opacity=0.3, rounded corners=5pt] 
    (2,-3.5) rectangle (7.5,-4.5);
    \fill[periwinkle, opacity=0.3, rounded corners=5pt] 
    (5,-0.5) rectangle (10.5,0.5);
    \fill[iceblue, opacity=0.3, rounded corners=5pt] 
    (11,-2.5) rectangle (12,0.5);
    \fill[periwinkle, opacity=0.3, rounded corners=5pt] 
    (9.5,-3.5) rectangle (12,-4.5);

    \foreach \y in {0,-2,-4} {
        \draw[thick] (2.5,\y) -- (11.5, \y);
    }

    \foreach \x in {2.5, 4,5.5,7,8.5,10,11.5} {
        \draw[thick,opacity=0.15] (\x,-2) -- (\x,-4);
        \draw[thick,opacity=0.15] (\x,0) -- (\x,-2);
    }
    
    \draw[thick, opacity=0.15] (2.5,-3) -- (11.5,-3);
    \draw[thick, opacity=0.15] (2.5,-1) -- (11.5,-1);

    \foreach \x in {2.5, 4,5.5,7,8.5,10,11.5} {
        \node[circle, fill=black!15] at (\x,-3) {};
        \node[circle, fill=black!15] at (\x,-1) {};
    }
    
    \draw[thick] (4,0) -- (4,-2);
    \draw[thick] (8.5,-2) -- (8.5,-4);
    \draw[thick] (11.5,-2) -- (11.5,0);
    
    \foreach \x in {2.5, 4,5.5,7,8.5,10,11.5} {
        \foreach \y in {0,-2,-4} {
            \node[circle, fill=pistacchio, draw=black] at (\x,\y) {};
        }
    }
\end{tikzpicture}
    \caption{Elementary building blocks for measurement-based quantum computing. $(a)$ We first consider the action of a single measurement on a single entangled state. $(b)$ Then, we concatenate entanglement and measurement to realize single-qudit gates on horizontal one-dimensional chains of arbitrary length. In addition, we require the measurement-based realization of an entangling gate between two qudits, which we achieve via transport through vertical edges using the configuration in $(c)$. Finally, we describe how to disassemble a two-dimensional resource state lattice into the previous building blocks via vertex deletion or the creation of new edges. Here, each horizontal line corresponds to the processing of a single qudit. For instance, via vertex deletion, one-dimensional chains of different lengths are cut out and used to implement single-qudit gates, while vertical edge creation allows the realization of entangling gates between two qudits in $(d)$.
    }
    \label{fig:setting}
\end{figure*}
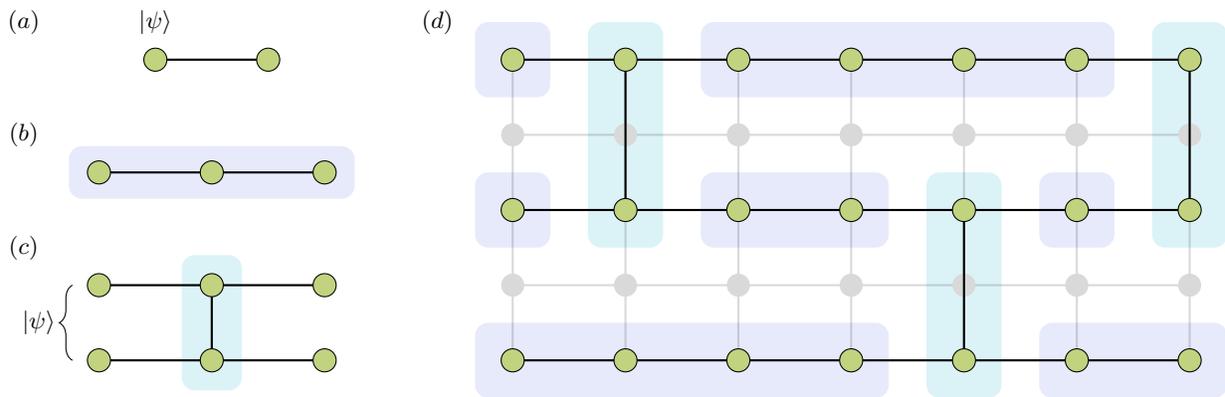

Qudits are high-dimensional quantum systems, native in many physical realizations of quantum information carriers such as energy levels of neutral atoms \cite{atomic_qudit,nuclearspin_qudits,ultracold_neutral_atoms} or trapped ions \cite{qudit_simulation_latticegaugetheories,nativequditentanglinggate,qudit_trappedions}, superconducting \cite{transmon_qudits,superconducting_qudit,transmon_qutrits}, and photonic systems \cite{photonic_qudit_d25,photonicqudit-32dimensions-shoralgorithm, photon_temporalmodes,photons_orbitalangularmomenta,qudit_siliconphotinicchip}.
Compared to their binary counterpart, leveraging qudits may reduce quantum circuit complexity for quantum computing \cite{quditcircuitcompression,QuditComputing, effiecentalgorithmswithqudits, QutritImprovementQuantumCircuit}, improve the security of quantum communication \cite{QuditQKD,QuditQKD_moreSecurity} or the error threshold of quantum codes \cite{EnhancedFaultTolerantComputing, ErrorThresholdsBitFlip, QuditSurfaceCode}, and enhance the performance of quantum metrology \cite{QuantumMetrolgyTransmon,ImprovedSensing} and quantum simulation \cite{qudit_simulation_latticegaugetheories,quditsimulation_fermionicsystems}.
Despite these theoretical advantages and emerging qudit hardware, the qudit approach to quantum information processing remains less developed and understood than the conventional qubit framework. As a result, much of current quantum information processing is artificially constrained to two-level systems.

While most previous work on quantum computation focuses on qubits and gate-based models, measurement-based quantum computing \cite{OnewayQC,Raussendorf_2003,MBQCwithDur,jozsa2005introductionmeasurementbasedquantum} has emerged as a viable alternative.
There, quantum information processing takes place solely through single-qubit measurements performed on a universal, entangled resource state, e.g., a cluster state \cite{ClusterStates} or, more generally, graph states \cite{Hein_2004,hein2006entanglementgraphstatesapplications}. This shifts the requirements for obtaining universal quantum computation from implementing sequences of universal single-qubit and two-qubit gates to the preparation of a single universal resource state together with local measurements. The measurement-based quantum computing paradigm has been extended to qudits using cluster or graph state resources \cite{Zhou_2003, Clark_2006, Booth_2023}.

Here, we consider alternative qudit resources with the same underlying entanglement topology corresponding to two-dimensional lattices but generated by different kinds of entangling gates. This small difference turns out to have far-reaching consequences in the qudit case, where not only the gates leading to suitable resource states are limited, but also the intrinsic complexity to process quantum information via measurements strongly depends on the choice of gate. In particular, for prime-power-dimensional qudits, we derive an upper bound on the maximum number of measurements to implement arbitrary single-qudit gates. This bound suggests that the cluster state is only optimal in even prime-power dimensions (in the sense of admitting self-inverse intrinsic gates), so that alternative qudit resource states could lead to improved computational efficiency and reduced resource overheads.

The alternative qudit resource states that we introduce are created by applying diagonal or block-diagonal Clifford entangling gates plus potentially non-standard initialization of the resource qudits, granting more experimental freedom in the preparation of a universal resource state.
For instance, in existing trapped-ion setups, a native two-qudit gate is available, the so-called light-shift gate, which has been used for the experimental generation of two-qudit entanglement \cite{nativequditentanglinggate}.

For these alternative resource states, we identify simple conversion instructions from the more prevalent quantum circuit picture to the measurement-based approach.
At the heart of this is an intrinsic gate associated with each resource state and determined by the entangling gate used for state preparation.
Furthermore, we prove a condition for the computational universality of the realizable gate set and describe how to render the computation deterministic despite the randomness of measurement outcomes.

For prime-power-dimensional qudits, we additionally find an upper bound on the number of measurements to realize an arbitrary single-qudit unitary that scales linearly in the dimension and the Pauli order of the intrinsic gate. Here, the intrinsic gate raised to the Pauli order corresponds to the identity up to a Pauli operator. Thus, our upper bound links different intrinsic operations to distinct computational overheads, indicating the suboptimality of the cluster state in odd prime-power dimensions and highlighting the potential of alternative qudit resources to improve computational efficiency.

In particular, we find that further optimal resources with self-inverse intrinsic gates (so minimal Pauli order, resulting in minimal measurement overhead) exist in prime-power dimensions with an even integer exponent. These resources can be realized via diagonal entangling interactions and cannot be mimicked by the qudit cluster state via adjusting measurement bases.

Furthermore, we demonstrate that Clifford circuits can be implemented in a single time step, as for qudit cluster state resources. Finally, we describe the required two-dimensional geometry of the resource stabilizer state that allows the implementation of any quantum circuit via single-qudit measurements only.

To summarize, our main results are as follows. First, we introduce a new class of universal qudit resource states characterized by different entangling gates. Second, we estimate the overhead for the measurement-based implementation of arbitrary single-qudit unitaries. Third, we discuss which of the considered resource states can be understood via qudit graph states with adjusted measurement bases and which cannot be translated in this way. As examples, we present explicit qutrit resource states, highlighting how the qutrit cluster state can be effectively adapted to achieve higher computational efficiency. Furthermore, we show the existence of optimal resources in prime-power dimensions with even exponent, which cannot be mimicked by a qudit graph state.
Lastly, we show that Clifford circuits can be realized in a single time step for these stabilizer state resources.

The paper is organized as follows. In Sec. \ref{sec:setting-methods}, we discuss the general setting of measurement-based quantum computing, in Sec. \ref{sec:background} we introduce the relevant mathematical framework to describe qudits, in Sec. \ref{sec:stabilizer-state-resources} we describe our generalized qudit stabilizer state resources, compare several universal resource states in Sec. \ref{sec:resource-comparison}, and, finally, conclude in Sec. \ref{sec:conclusion}.

\section{Setting and methods}
\label{sec:setting-methods}

Measurement-based quantum computing \cite{OnewayQC,Raussendorf_2003,MBQCwithDur,jozsa2005introductionmeasurementbasedquantum, Zhou_2003, Clark_2006, Booth_2023} relies on an entangled resource state, that is manipulated solely via single-system measurements to process quantum information.
In the standard approach to measurement-based quantum computing, a cluster state \cite{ClusterStates,Zhou_2003} is used, a graph state \cite{Hein_2004, hein2006entanglementgraphstatesapplications} with a regular square lattice structure.

We visualize resources via graphs, where qudits are associated with vertices and edges with the application of an entangling gate. The entangling gate and the initial state of the qudits that characterize our qudit stabilizer state resources are varied later. Here, the entanglement arises either via a natural interaction of the qudits or via explicitly preparing the resources, so applying the gate.

To understand the correspondence between single-qudit measurements and quantum information processing, we always start by considering the action of a single measurement on a single two-qudit entangled state, as shown in Fig. \ref{fig:setting} $(a)$. As discussed later, only certain types of single-qudit measurements are permitted to guarantee unitary quantum information processing. Therefore, the class of single-qudit gates being implemented with a single measurement is restricted, in particular, the so-called intrinsic gate of the resource is always realized. This intrinsic gate is determined by the entangling gate and the initial state of the resource qudits.

Repeated single-qudit measurements on one-dimensional resource chains then allow implementing a sequence of single-qudit gates, shown in Fig. \ref{fig:setting} $(b)$. Each measurement has a random outcome, however, all possible outcomes can be related to the desired outcome and accounted for by adapting the single-qudit measurement bases on subsequent turns and post-processing. As a part of the universality proof for our generalized qudit resources, we later translate arbitrary single-qudit gates into measurement patterns.

Next, we require the ability to perform entangling gates measurement-based, for which a resource configuration shown in Fig. \ref{fig:setting} $(c)$ is used. Here, an arbitrary two-qudit input is transported along its respective one-dimensional horizontal chain while the vertical edge realizes an entangling gate. The ability to perform any entangling gate suffices for achieving a universal gate set \cite{MathematicsQC,CriteriaQuditUniversality,QuditComputing,proctor2019quantuminformationgeneralquantum}.

Finally, we need the ability to combine these elementary building blocks at will, so one-dimensional chains as in Fig. \ref{fig:setting} (b) of arbitrary length for single-qudit gate implementation and the shape in Fig. \ref{fig:setting} (c) to realize entangling gates measurement-based. For this, we consider resource states structured as two-dimensional lattices, Fig. \ref{fig:setting} $(d)$, and identify single-qudit measurements that achieve either vertex deletion or edge creation to cut out the necessary building blocks.

In the next section, we discuss the details of these steps for the standard cluster state resource before introducing generalized qudit stabilizer state resources in Sec. \ref{sec:stabilizer-state-resources}.

\section{Theoretical background}
\label{sec:background}

The standard resource for measurement-based quantum computing is a cluster state or, more generally, certain graph states \cite{Hein_2004, hein2006entanglementgraphstatesapplications}. Measurement-based quantum computing with cluster state qudits of arbitrary finite dimension was first introduced in Ref. \cite{Zhou_2003}. Moreover, qudit resources based on matrix product states, typical for natural many-body ground states, have been studied \cite{MPSfinitebyproductgroup, PhysRevLett.98.220503,MPSglobalsymmetry}, and finite group cluster states, created by controlled group multiplication gates \cite{GeneralizedClusterStateFiniteGroup}.

We introduce the integer-ring and finite-field descriptions of qudits and the respective working mechanism of cluster state measurement-based quantum computing in Secs. \ref{sec:MBQC-integer-ring} and \ref{sec:MBQC-finite-field}. Then, we describe the computation via single-qudit measurements in terms of what we call an intrinsic gate, the Hadamard gate for the cluster state. Our generalized qudit stabilizer state resources will then have other intrinsic gates associated with them.

\subsection{Measurement-based quantum computing with integer-ring qudits}
\label{sec:MBQC-integer-ring}

\subsubsection{The integer-ring Pauli and Clifford groups}

For any finite dimension $d$, one can label the qudit basis states of a $d$-dimensional Hilbert space in terms of the integer ring $\mathbb{Z}_d = \{0,\hdots ,d-1 \}$, where addition is performed modulo $d$ \cite{QuditComputing,LinearizedStabilizerFormalism}.
We identify the basis states of an integer-ring qudit with the eigenstates of the generalized Pauli operator $Z_d$, often called clock operator, and denote them via
\begin{equation}
    \{ \ket{j}\}_{j=0}^{d-1} = \{ \ket{j_Z}\}_{j=0}^{d-1},
\end{equation}
where $Z_d \ket{j} = (\omega_{d})^j \ket{j}$ with $\omega_{d} = e^{2\pi i/d}$ being the $d$-th root of unity. If $d$ is prime, the integer ring $\mathbb{Z}_d$ becomes a field and coincides with a finite-field qudit introduced in the next section.

The generalized Pauli $X_d$ operator acts on computational basis states via a shift, $ X_d \ket{j} = \ket{j+1}$. To obtain a generalized $Y_d$ operator that has order $d$ and satisfies $X_d Y_d Z_d = \tau_d I_d$ with the $d \cross d$ identity matrix $I_d$ and $\tau_d^2 = e^{\frac{2 \pi i }{d}}$ (to generalize the relationship $XYZ = i I_2$ in the case of qubits), we can choose $\tau_d = (-1)^d e^{\frac{\pi i}{d}}$ and set $Y_d = \tau_d X_d^\dagger Z_d^\dagger = \tau_d X_d^{-1} Z_d^{-1}$ \cite{LinearizedStabilizerFormalism}. This non-trivial choice of $\tau_d$ that differs for the even and odd dimensional cases is necessary to ensure that $Y_d$ has order $d$ since otherwise for even-dimensional $d$, we would have $(X_d^\dagger Z_d^\dagger)^d = - I_d$.

The single-qudit Pauli group $\mathcal{P}$ is then generated by
\begin{equation}
    \mathcal{P} = \langle X_d, Y_d, Z_d \rangle = \langle \tau_d I_d, X_d, Z_d \rangle
\end{equation}
and the generalized Pauli operators satisfy the commutation relation \cite{LinearizedStabilizerFormalism}
\begin{equation}
    X_d^b Z_d^a = \tau_d^{-2 ab} Z_d^a X_d^b = \omega_d^{-ab}  Z_d^a X_d^b.
\label{eq:commutation-modular-Paulis}
\end{equation}

Stabilizer states $\ket{\psi}$ are uniquely characterized via an associated stabilizer group $\mathcal{S}$, an abelian (commutative) subgroup of the Pauli group, via $s \ket{\psi} = \ket{\psi}$ for all $s \in \mathcal{S}$. Typically, we describe the stabilizer group $\mathcal{S} = \langle g_1, \hdots, g_l \rangle$, via a minimal generating set $\{g_1, \hdots, g_l \} \in \mathcal{P}^{\otimes l}$.
For prime dimensions $d$, $n$ generators suffice to describe an $n$-qudit stabilizer state, whereas for composite $d$, the number of generators $l$ may be strictly larger than $n$, however, it is upper bounded by $2n$ \cite{LinearizedStabilizerFormalism}. For instance, the single-ququart state $\frac{1}{\sqrt{2}}(\ket{0}+\ket{2})$ is a stabilizer state with minimal generating set $\{ X^2, Z^2 \}$.

The integer-ring single-qudit Clifford group that leaves the Pauli group $\mathcal{P}$ invariant under conjugation is generated by $Z_d$ and the operators \cite{Farinholt_2014}
\begin{equation}
    H_d = \frac{1}{\sqrt{d}} \sum_{j,k = 0}^{d-1} \omega_{d}^{kj} \ket{j} \bra{k}, \quad S_d = \sum_{j =0 }^{d-1} \tau_d^{j^2} \ket{j} \bra{j}
\end{equation}
with the Hadamard gate $H_d$ and the phase gate $S_d$.
Due to $\tau_d$ differing for even and odd dimension $d$, the phase gate is often defined differently for odd and even dimensional qudits instead of introducing $\tau_d$ \cite{BenchmarkingCliffordgates,proctor2019quantuminformationgeneralquantum} 
(sometimes also defining the $Y_d$ gate differently \cite{proctor2019quantuminformationgeneralquantum}).
Some relevant Pauli conjugation relations are presented in Appendix \ref{app:integerring-cliffords}.
Supplementing the single-qudit Clifford group with the controlled-$Z_d$ gate
\begin{equation}
    CZ_d = \sum_{k,j=0}^{d-1} \omega_d^{kj} \ket{kj} \bra{kj} 
\end{equation}
or $CX_d= \sum_{k,j} \ket{k} \bra{k} \otimes \ket{j+k} \bra{j}$ we generate the $n$-qudit integer-ring Clifford group of any finite dimension $d$ \cite{Farinholt_2014,QuditsArbitraryDimension,BenchmarkingCliffordgates,proctor2019quantuminformationgeneralquantum}.

Note that the generalized $X_d$, $Y_d$, and $Z_d$ gates are no longer self-adjoint, so they do not have real eigenvalues and are not observables anymore. However, their respective eigenstates still form an orthonormal basis of the qudit Hilbert space.
In particular, the eigenvectors of $X_d$ are given by
\begin{equation}
    \ket{k_X} = H_d \ket{k_Z} = H_d X_d^k \ket{0_Z} = Z_d^k H_d \ket{0_Z} = Z_d^k \ket{0_X},
    \label{eq:relating-outcome-to-desired}
\end{equation}
and the operator $Y_d$ has the eigenstates $\ket{k_Y} = S_d \ket{k_X} = S_d H_d \ket{k_Z}$, see Appendix \ref{app:integerring-cliffords}.

Identifying a qudit with a vertex $v$ and the application of a $(CZ_d)^{w_{uv}}$ gate with an edge of weight $w_{vu}$ between the corresponding vertices $u$ and $v$, initialized in
\begin{equation}
    \ket{+} \coloneq \ket{0_X} = \frac{1}{\sqrt{d}} \sum_k \ket{k},
\end{equation}
a graph state has stabilizer group generators
\begin{equation}
    g_v = X_v \prod_{u \in N_v} Z_u^{w_{uv}}
\end{equation}
for each vertex $v$, where $N_v$ denotes the neighborhood of $v$.
This can be seen by conjugating the stabilizer of $\ket{+}$, namely $X_v$ (with identities elsewhere), by $CZ_d$.
Therefore, one can use graphs, such as in Fig. \ref{fig:setting}, to visualize cluster state resources, where $w_{uv} = 1$ for all edges.

\subsubsection{Computation via single-qudit measurements}

To understand how single-qudit measurements implement quantum gates, we start with the scenario in Fig. \ref{fig:setting} $(a)$ for the cluster state, where edges correspond to the application of the $CZ_d$ gate and all resource qudits are initialized in the equal superposition state $\ket{+}$.

Applying the $C Z_d$ gate to an arbitrary qudit input state $\ket{\psi} = \sum_{l=0}^{d-1} \alpha_l \ket{l}$ with $\alpha_l \in \mathbb{C}$ and $\sum_{l=0}^{d-1} |\alpha_l|^2 = 1$ and one qudit in $\ket{+}$, yields the total state
\begin{equation}
    CZ_d (\ket{\psi}_1 \ket{+}_2) = \sum_{l,m=0}^{d-1} \alpha_l \omega_{d}^{lm} \ket{l}_1 \ket{m}_2 = \sum_{l=0}^{d-1} \alpha_l \ket{l}_1 H_d \ket{l}_2.
\end{equation}

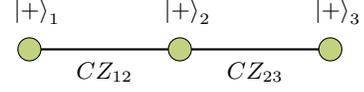
\begin{figure}[t]
    \centering
    \begin{tikzpicture}
    
    \coordinate (1) at (-2,0);  
    \coordinate (2) at (0,0); 
    \coordinate (3) at (2,0);

    \draw[thick] (1) -- (3);

    \foreach \i in {1,2,3} {
        \node[circle, fill=pistacchio, draw=black] at (\i) {};
    }
    \node at (-1,-0.3) {$CZ_{12}$};
    \node at (1,-0.3) {$CZ_{23}$};
    \node at (-1.9,0.5) {$\ket{+}_1$};
    \node at (0.1,0.5) {$\ket{+}_2$};
    \node at (2.1,0.5) {$\ket{+}_3$};
\end{tikzpicture}
    \caption{Concatenation of the two-qudit protocol for the measurement-based implementation of a single-qudit unitary. The first qudit $\ket{+}_1$ acts as an input state to be processed and is connected with a controlled-phase gate $CZ_{12}$ to {$\ket{+}_2$} and subsequently measured in the basis $\{D_{\Vec{\phi}} \ket{k_X} \}_k$. Repeating this procedure, so entangling the second qudit, which is in the state $\ket{\psi} \coloneq H_d D_{\Vec{\phi}} Z^{-k} \ket{+}$ after the first measurement with outcome $k$, via $CZ_{23}$ with the third qudit $\ket{+}_3$ and measuring the second qudit subsequently is equivalent to first applying both entangling $CZ$ operations for preparing the three-qudit linear cluster state resource and afterwards performing both single-qudit measurements due to the commutation of all involved operations. }
    \label{fig:concetenation-two-qubit-protocol}
\end{figure}

Not every single-qudit measurement on the first qudit of $CZ_d (\ket{\psi} \ket{+})$ induces transport of quantum information with simultaneous unitary processing. Considering any possible single-qudit measurement outcome $\ket{m} = \sum_j m_j \ket{j_Z}$ with normalized $\sum_j |m_j|^2 = 1$, the posterior state for this result becomes
\begin{equation}
    \begin{aligned}
        & \bra{m}_1 \sum_{k=0}^{d-1} \alpha_k \ket{k_Z}_1  H_d \ket{k_Z}_2
        \propto \sum_{k=0}^{d-1} \alpha_k m_k^*  H_d \ket{k_Z}_2,
    \end{aligned}
\end{equation}
where we need to normalize the resulting single-qudit state with $\sqrt{\sum_{j} |\alpha_j|^2 |m_j|^2}$, the probability to observe the result $\ket{m}$. To describe the impact of the single-qudit measurement in terms of a unitary transformation
    \begin{equation}
    D(\Vec{m}) \coloneq \frac{1}{\sqrt{\sum_{j} |\alpha_j|^2 |m_j|^2  } }
    \left( \begin{array}{ccc}
        m_0^* &  \\
         & \ddots \\
         & & m_{d-1}^*
    \end{array} \right)
\end{equation}
the modulus $|m_j|$ has to be constant since unitarity of $D(\Vec{m})$ requires that
\begin{equation}
    \frac{|m_k|^2}{\sum_{j} |\alpha_j|^2 |m_j|^2} = 1 \quad \forall k \in \{0,...,d-1\}.
\end{equation}
Hence, the only allowed measurement bases consist of equal superposition states. Furthermore, the only unitary gate that we can execute with a single-qudit measurement is $H_d D_{\Vec{\phi}}$.
In particular, the Hadamard gate $H_d$ is always implemented for any choice of $D_{\Vec{\phi}}$, where $D_{\Vec{\phi}} = \textnormal{diag}(e^{i \phi_1}, \hdots, e^{i \phi_d})$, so that we refer to it as the intrinsic gate of the cluster state resource. More precisely, the intrinsic gate specific to a resource state corresponds to the single-qudit operation that is implemented if one performs an $X$ basis measurement at zero angle. If we have a qudit graph state with equal edge weight $w$, the intrinsic gate becomes $H M(w)$.

An appropriate basis for unitary evolution is the $X_d$ basis, rotated by $D_{\Vec{\phi}}$, so
\begin{equation}
    \{ D_{\Vec{\phi}} \ket{k_X} \}_k = \{ D_{\Vec{\phi}} Z^k \ket{0_X} \}_k.
\end{equation}
Here, we use that according to Eq. \eqref{eq:relating-outcome-to-desired}, a non-zero outcome $\ket{k_X}$ of a measurement in the $X_d$ basis can always be related to the outcome $\ket{0_X}$, the desired outcome, via the $Z_d$ operator.
Measuring the first qudit in this basis with outcome $k \in \mathbb{Z}_d$, we effectively move the quantum information $\ket{\psi}$ to the second qudit while simultaneously processing it with $ H_d Z^{-k} D_{\Vec{\phi}}^\dagger = X^k H_d D_{\Vec{\phi}}^\dagger $ due to
\begin{equation}
\begin{aligned}
     \bra{0_X}_1 Z^{-k} D_{\Vec{\phi}}^\dagger \sum_{l } \alpha_l  \ket{l}_1 H_d  \ket{l}_2 \propto H_d Z^{-k}  D_{\Vec{\phi}}^\dagger \ket{\psi}_2.
\end{aligned}
\end{equation}
For qubits, the gate $D_{\Vec{\phi}}$ corresponds to a $Z$ rotation $R_Z(\phi) = e^{-i \phi Z}$, where $Z \equiv Z_2$, so that $D_{\Vec{\phi}}$ can be understood as a generalization of $R_Z(\phi)$ to qudits.

We concatenate this two-qudit protocol, see Fig. \ref{fig:concetenation-two-qubit-protocol}. This is possible since we have considered an arbitrary single-qudit state $\ket{\psi}$ as the input of the two-qudit protocol. Furthermore, the measurement of the first qudit and the controlled-$Z_d$ gate that entangles the next two qudits act on disjoint sets of qudits, so that we can reorder the operations and first create our resource state. This corresponds to the one-dimensional resource chains in Fig. \ref{fig:setting} $(b)$.

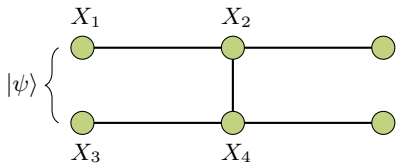
\begin{figure}[t]
    \centering
    \begin{tikzpicture}

    \coordinate (1) at (-2,0); 
    \coordinate (2) at (0,0); 
    \coordinate (3) at (2,0); 

    \coordinate (4) at (-2,-1); 
    \coordinate (5) at (0,-1); 
    \coordinate (6) at (2,-1);

    \draw[thick] (1) -- (2) -- (3);
    \draw[thick] (4) -- (5) -- (6);
    \draw[thick] (2) -- (5);

    \foreach \i in {1,2,3,4,5,6} {
        \node[circle, fill=pistacchio, draw=black] at (\i) {};
    }
    
    \draw [decorate,decoration={brace,amplitude=5pt,mirror,raise=9pt}]
    (1) -- (4) node [black,midway,xshift=-0.8cm] 
    {$\ket{\psi}$};
    
    \node at (-1.95,0.4) {$X_1$};
    \node at (0.05,0.4) {$X_2$};
    \node at (-1.95,-1.4) {$X_3$};
    \node at (0.05,-1.4) {$X_4$};
\end{tikzpicture}
    \caption{Realization of a measurement-based $CZ_d$ gate
    acting on the two-qudit input $\ket{\psi}$ via a measurement pattern on a six-qudit cluster state, using an existing $CZ_d$ connection \cite{Zhou_2003}.}
    \label{fig:imprimitive-two-qudit}
\end{figure}

Depending on the measurement outcomes, we have undesired $Z^{-k}$ operators intertwined in the operations $H_d  D_{\Vec{\phi}}$ that we want to realize. These random by-product operators can be moved to the end of the computation being performed and classically post-processed. More precisely, since the Hadamard gate is Clifford, it allows the propagation of any Pauli operator by conjugation. Diagonal gates $D_{\Vec{\phi}}$ commute with $Z_d$ and are conjugated by $X_d$ to a new diagonal gate $D_{\Vec{\phi}'} = X^k D_{\Vec{\phi}} X^{-k}$, where $\Vec{\phi}'$ has permuted entries compared to $\Vec{\phi}$.
Hence, in general, we need to adjust our generalized $Z_d$ rotation $D_{\Vec{\phi}}$ adaptively to another diagonal gate $D_{\Vec{\phi}'}$ depending on previous measurement outcomes analogous to how in the qubit case the measurement angle $\phi$ in $R_Z(\phi)$ may need to be adjusted to $-\phi$.
Then, the Pauli by-product $(X_d)^{x} (Z_d)^{z}$ can be accounted for via post-processing since the $Z_d$ operator has no effect on measurements in the computational basis of the output, and the $X_d$ gate leads to re-interpreting the measurement outcome.

A measurement in the $Y_d$ basis realizes $H_d S_d^\dagger$, so that we can implement Clifford circuits solely via $X_d$ and $Y_d$ measurements, as for qubits. Here, no adaptation of diagonal gates (which are themselves Clifford, namely $S_d$) is required, and we can perform all measurements simultaneously in a single time step. To remove the Hadamard gate in $H_d S_d^\dagger$, one can concatenate the $Y_d$ measurement with three $X_d$ measurements. Then, using that $H_d^4 = I_d$ for all finite dimensions $d$ (see Appendix \ref{app:integerring-cliffords}), one obtains $S_d^{\dagger}$ up to a Pauli by-product.

Universal quantum computing with qudits requires the implementation of arbitrary single-qudit unitaries, so also of non-Clifford gates. Conceptually simple exact universal gate sets can be constructed from the Hadamard gate and diagonal phase gates \cite{proctor2019quantuminformationgeneralquantum, Zhou_2003}.
This gate set is naturally realized via single-qudit measurements on the cluster state and such a set, along with any two-qudit entangling gate, is then sufficient for exact universal quantum computing \cite{MathematicsQC,CriteriaQuditUniversality,QuditComputing,proctor2019quantuminformationgeneralquantum}. Later on, we explicitly decompose arbitrary single-qudit unitaries into measurement patterns that can be realized on one-dimensional qudit stabilizer state resource chains, providing, in addition, overhead estimates associated with different resources.

The easiest way to realize a measurement-based entangling two-qudit gate uses an existing edge of the cluster state, as shown in Figs. \ref{fig:setting} $(c)$ and \ref{fig:imprimitive-two-qudit} \cite{Zhou_2003}. Transporting a two-qudit input through the edge is then equivalent to the $CZ_d$ gate being applied to both qudits of the input after the $X_1$ and $X_2$ and before the $X_3$ and $X_4$ measurements.

These elementary building blocks, i.e., one-dimensional lines as well as the constellation in Fig. \ref{fig:imprimitive-two-qudit} can be cut out from a universal resource state via single-qudit measurements. More precisely, a measurement in the $Z_d$ basis removes a vertex together with all its edges. In turn, a measurement in the $Y_d$ basis removes a vertex from the graph while simultaneously inverting neighborhood relations (called local complementation), so that edges can be created \cite{hein2006entanglementgraphstatesapplications}.
Hence, we can disassemble the cluster state resource to combine arbitrary single-qudit gates with the entangling gate $CZ_d$ and realize a desired quantum circuit, as discussed in the previous Sec. \ref{sec:setting-methods}.

\subsection{Measurement-based quantum computing with finite-field qudits}
\label{sec:MBQC-finite-field}

In this section, we discuss the definition of the Pauli and Clifford groups for a mathematical description of prime-power-dimensional qudits that relies on finite fields instead of integer rings. Measurement-based quantum computing with such finite-field qudits then works analogously to the previous case replacing the definitions of the respective Pauli and Clifford operators. However, the finite-field structure provides more powerful mathematical tools. For instance, it enables an elegant and more efficient decomposition of single-qudit unitaries into measurement patterns, as demonstrated later.

\subsubsection{The finite-field Pauli group}

Qudits of prime-power dimension $d = p^m$ with $p$ being a prime, $m \in \mathbb{N}$, allow us to identify the computational basis states with finite-field elements
\begin{equation}
    \mathbb{F}_{p^m} \cong \mathbb{F}_{p}[\xi] / \langle f(\xi) \rangle
\end{equation}
instead of integer ring elements, where $\mathbb{F}_p[\xi]$ is a polynomial ring in the variable $\xi$ with coefficients from $\mathbb{F}_p = \mathbb{Z}_p$ and $f(\xi)$ is an irreducible polynomial (meaning that it cannot be factored into non-constant polynomials) of degree $m$ \cite{heinrich2021stabiliser,Vourdas_2007,beauregard2003quantumarithmeticgaloisfields}.

Hence, any finite-field qudit can be represented via a polynomial $\ket{a_0 + a_1 \xi + \hdots + a_{m-1} \xi^{m-1}}$ with $a_i \in \mathbb{F}_p$. Analogous to how in the integer ring $\mathbb{Z}_d$, computations are performed modulo $d$, in the finite field, computations are performed modulo the irreducible polynomial $f(\xi)$ and modulo $p$. In Appendix \ref{app:finite-field-ququarts}, we provide a practical example of finite-field ququarts, where $p=m=2$.

For $m >1$, we need to modify the definitions of Pauli and Clifford operators, such that they act on finite-field elements.

The finite-field Pauli gates are then defined via \cite{heinrich2021stabiliser}
\begin{equation}
    Z(z) \ket{u} = \chi(z \cdot u) \ket{u}, \quad X(x) \ket{u} = \ket{u + x},
\end{equation}
where $\chi(t) = \omega_p^{\tr(t)}$ with $\omega_p = e^{\frac{2 \pi i}{p}}$. Here, the finite-field trace $\tr(t)$ is the trace of the multiplication map with $t \in \mathbb{F}_{p^m}$ and it associates basis state polynomials with integers via
\begin{equation}
    \tr(t): \mathbb{F}_{p^m} \mapsto \mathbb{F}_{p} = \mathbb{Z}_p, \quad a \mapsto \tr(a) = \sum_{j=0}^{m-1} a^{p^j}.
\end{equation}
For $m=1$, so prime-dimensional qudits, the integer-ring and finite-field definitions of Pauli operators and, hence, the Clifford group, coincide with $Z(z)$ becoming $(Z_d)^z$ and $X(x)$ becoming $(X_d)^x$.
We have
\begin{equation}
\begin{aligned}
    & Z(z) X(x) = \sum_{u \in \mathbb{F}_{p^m}} \chi(z \cdot (u+x)) \ket{u+x} \bra{u}
    \\ & = \chi(z \cdot x) \sum_{u \in \mathbb{F}_{p^m}} \ket{u+x} \bra{u} \chi(z \cdot u)
     = \chi(z \cdot x) X(x) Z(z)
\end{aligned}
\label{eq:commutation-finitefield-Paulis}
\end{equation}
in analogy to Eq. \eqref{eq:commutation-modular-Paulis}.

In the odd-dimensional case with $p \neq 2$, the Heisenberg-Weyl group, the prime-power generalization of the Pauli group, consists of the Weyl operators \cite{heinrich2021stabiliser}
\begin{equation}
    W(z,x,t) = \chi(t) \chi(-2^{-1} z x) Z(z) X(x),
\end{equation}
whereas in the even case, $p = 2$, the Weyl operators are defined via
\begin{equation}
    W(z,x,t) = \chi_4(t) \chi_4(-z x) Z(z) X(x)
\end{equation}
with $\chi_4(t) = i^{\tr_4(t)}$. The trace $\tr_4(t):\mathbb{GR}(4,m) \mapsto \mathbb{Z}_4$ of the multiplication map is now computed in the Galois ring $\mathbb{GR}(4,m)$ with $4^m$ elements, so an extension of the $\mathbb{Z}_4$ ring with extension degree $m$. The group formed by the operators $W(z,x,t)$ is called the Heisenberg-Weyl group. In Appendix \ref{app:finite-field-ququarts}, we provide explicit expressions for finite-field ququarts as an example.

Stabilizer states of $n$ qudits can again be defined via abelian subgroups of the Heisenberg-Weyl group, where $n$ stabilizer group generators suffice \cite{heinrich2021stabiliser}.
Furthermore, in contrast to integer-ring qudits, any stabilizer state is locally Clifford equivalent to a graph state \cite{schlingemann2001stabilizercodesrealizedgraph,Grassl}. In particular, the stabilizer group generators of a finite-field qudit graph state correspond to
\begin{equation}
    g_v = X_v(x) \prod_{u \in N_v} Z_u(w_{uv} x)
\end{equation}
for all $x \in \mathbb{F}_d$, each vertex $v$, and edge weights $w_{uv} \in \mathbb{F}_d$ (with $w_{uv} = 1$ for the cluster state, such as in Fig. \ref{fig:setting}).

\subsubsection{The finite-field Clifford group}
\label{subsec:finite-field-Clifford}

We define the finite-field Clifford group as a structure-preserving normaliser of the Heisenberg-Weyl group, i.e., as preserving the commutation relations in Eq. \eqref{eq:commutation-finitefield-Paulis} and, in addition, being linear in the argument $(z,x)$ of Pauli operators $Z(z) X(x)$. The full unitary normaliser of the Pauli group is larger and consists of Clifford transformations acting on $m$ many $p$-dimensional qudits \cite{heinrich2021stabiliser}.

The explicit set of generators of the Clifford group depends on the characteristic $p$. The representation in the case of an odd prime $p$ relies on the characteristic not being two, using $\chi(2^{-1}t) = \omega^{2^{-1}\tr(t)}$ during the construction of the Clifford group generators. The inverse of two does not exist for $p=2$, and to mimic the behavior of the square root of $\omega$, one needs to introduce a character $\chi_4$ of an extension ring over $\mathbb{Z}_4$ in the even case \cite{heinrich2021stabiliser}. This is similar to the previous introduction of $\tau_d$ as one of the two square roots of $\omega$ depending on whether the dimension is even or odd for the integer-ring Clifford group.
Supplementing any single-qudit non-Clifford gate renders the Clifford group universal \cite{heinrich2021stabiliser}. We later explicitly decompose arbitrary single-qudit unitaries into operators from the measurement-based gate set to show exact universality.

The finite-field Clifford group \cite{heinrich2021stabiliser} is generated by $Z(1)$, $ X(1)$, the multiplication gates
\begin{equation}
    M(\lambda) = \sum_{x \in \mathbb{F}_{p^m}} \ket{\lambda x} \bra{x},
\end{equation}
for $0 \neq \lambda \in \mathbb{F}_d$, the $CX = \sum_{(j,k) \in (\mathbb{F}_{d})^2} \ket{j} \bra{j} \otimes \ket{j+k}  \bra{k}$ or $CZ = \sum_{j,k} \chi(jk) \ket{j} \bra{j} \otimes \ket{k} \bra{k}$ gate, the finite-field Hadamard gate
\begin{equation}
    H_d^F = \frac{1}{\sqrt{d}} \sum_{u,v \in \mathbb{F}_{d}} \chi(u v) \ket{u} \bra{v},
\end{equation}
which is equal to $H_d$ for prime-dimensional qudits, and the finite-field phase gate $S^F(\lambda)$. The latter is defined as
\begin{equation}
    S_d^F(\lambda) = \sum_{x \in \mathbb{F}_{p^m}} \chi(2^{-1} \lambda x^2) \ket{x} \bra{x}
    \label{eq:finitefield-phasegate-odd}
\end{equation}
for $p \neq 2$ and for $p=2$ as
\begin{equation}
    S_d^F = \sum_{x \in \mathbb{F}_{2^m}} \chi_4(x^2) \ket{x} \bra{x}.
    \label{eq:finitefield-phasegate-even}
\end{equation}
Note that for $d=2$, we have that $S_2^F = S$, and for odd prime dimensions $d=p$, it is $S_d^F(1) = S_d$, so that the finite-field definition coincides with the integer-ring definition (see Appendix \ref{app:conjugation-finitefield-cliffords}).

As an explicit example, we consider the smallest-dimensional finite-field qudits, namely ququarts, in Appendix \ref{app:finite-field-ququarts}, explicitly constructing the matrix representation of the $S_4^F$ gate. Additionally, we provide some Pauli conjugation relations for finite-field Clifford gates in Appendix \ref{app:conjugation-finitefield-cliffords} such as
\begin{equation}
    H_d^F X(k) (H_d^F)^\dagger = Z(k), \quad
    H_d^F Z(k) (H_d^F)^\dagger = X(-k)
\end{equation}
in analogy to the integer-ring Hadamard $H_d$ conjugation relation.

As for integer-ring qudits, we can then relate a non-zero $X$ measurement outcome to the desired outcome $\ket{0_X}$ via
\begin{equation}
    \begin{aligned}
        & \ket{k_X} = H_d^F X(k) \ket{0_Z}
         = Z(k) H_d^F \ket{0_Z} = Z(k) \ket{0_X}.
    \end{aligned}
\end{equation}
The Pauli by-product propagation is achieved in the same way as for integer-ring qudits using that, for a diagonal gate $D_{\Vec{\phi}}$, $X(k) D_{\Vec{\phi}} X(-k)$ remains diagonal.

\section{Qudit stabilizer state resources}
\label{sec:stabilizer-state-resources}

In the following, we consider qudit stabilizer state resources for measurement-based quantum computing which are not characterized by controlled-phase gates but instead by a class of more general entangling gates, starting with diagonal Clifford gates in Sec. \ref{sec:diagonal-entangling-gates} and then moving to block-diagonal Clifford gates in Sec. \ref{sec:blockdiagonal-entangling-gate}. Computation by single-qudit measurements is then understood as relying on an intrinsic single-qudit Clifford gate associated with the resource state; for the cluster state, this is the Hadamard gate.
For both classes of generalized resource states, we provide an example of a qutrit resource state allowing for more efficient quantum information processing with the qutrit cluster state by adjusting measurement bases.

\subsection{Diagonal Clifford entangling gates}
\label{sec:diagonal-entangling-gates}

\subsubsection{Intrinsic gate associated with resource}  

Commuting gates, such as diagonal gates, have the advantage that we do not need to keep track of the temporal order in which the entangling gates act on the resource qudits. Thus, these resources can be visualized with graphs, where qudits correspond to vertices and edges to the application of the entangling gate.

To determine an intrinsic gate that drives computation upon measurements on one-dimensional resource chains, we consider the two-qudit scenario of Fig. \ref{fig:setting} $(a)$.
Diagonal two-qudit entangling gates $G_E$ apply different phases $\theta_{jk}$ on states $\ket{jk}$ with $j,k \in \{ 0,...,d-1\} = \mathbb{Z}_d$ (or in $\mathbb{F}_d$ for finite-field qudits),
\begin{equation}
    G_E \coloneq \sum_{j,k=0}^{d-1} e^{i \theta_{jk}} \ket{jk} \bra{jk}.
    \label{eq:diagonal-entangling}
\end{equation}
Both the controlled-phase gate and the light-shift gate, a qudit entangling gate used in existing trapped-ion setups \cite{nativequditentanglinggate}, are special cases of such entangling gates.

Applying the diagonal entangling gate to a qudit pair, $G_E \ket{+} \ket{+}$, the reduced density matrix becomes
\begin{equation}
\begin{aligned}
    \rho &= \frac{1}{d^2} \sum_{a,j,l} e^{i(\theta_{ja}- \theta_{la})} \ket{j} \bra{l} 
    \\ &= \frac{1}{d} \left( \begin{array}{cccc}
         1& c_{0,1} & \hdots & c_{0,d-1} \\
         c_{1,0} & \ddots & & \vdots \\
         \vdots & & \ddots & c_{d-2,d-1} \\
        c_{d-1,0} & \hdots & c_{d-1,d-2} & 1
    \end{array} \right),
\end{aligned}
\end{equation}
where
\begin{equation}
    c_{j,k} \coloneq \frac{1}{d} \sum_{a=0}^{d-1} e^{i (\theta_{ja}- \theta_{ka} )}, \textnormal{ and } c_{j,k} = c_{k,j}^*.
    \label{eq:off-diagonal-reduced-density-matrix}
\end{equation}

For the two-qudit state $G_E \ket{+} \ket{+}$ to be maximally entangled, the single-qudit reduced state has to be maximally mixed, so all eigenvalues have to be $1/d$. We are interested in preparing maximally entangled two-qudit states to transport arbitrary quantum states via single-qudit measurements; for instance, a single qudit that is part of a maximally entangled multi-qudit state can only be faithfully transported if the two-qudit resource state itself is maximally entangled.

We want to understand the constraints on the angles $\theta_{jk}$ that yield $\rho = \frac{1}{d} I$, so $c_{j,k} = \delta_{j,k}$ with $\delta_{j,k}$ being the Kronecker delta. Fortunately, we do not need to explicitly solve $c_{j,k} = \delta_{j,k}$. Instead, these conditions are the exact requirements for unitary evolution of the quantum information upon single-qudit measurements, as we argue in the following.

Given the entangling gate $G_E$, we can rewrite
\begin{equation}
    G_E \ket{+} \ket{+} = \frac{1}{d} \sum_{j,k=0}^{d-1} e^{i \theta_{jk}} \ket{j} \ket{k} =  \frac{1}{d} \sum_{j,k=0}^{d-1} \ket{j} G_j \ket{k},
\end{equation}
introducing the single-qudit phase gates
\begin{equation}
     G_j \ket{k} \coloneq e^{i \theta_{jk}} \ket{k}.
\end{equation}
Applying $G_E$ on two qudits initialized in $\ket{+}$ yields
\begin{equation}
\begin{aligned}
    & G_E \ket{+} \ket{+} = \frac{1}{\sqrt{d}} \sum_{j= 0}^{d-1} \ket{j} G_j \ket{0_X} = \frac{1}{\sqrt{d}} \sum_{j,k= 0}^{d-1} \ket{j} G_k \ket{0_X} \delta_{j,k} \\ & = \frac{1}{\sqrt{d}} \sum_{j=0}^{d-1} \ket{j} \left(\sum_{k=0}^{d-1} G_k \ket{0_X} \bra{k_Z} \right) \ket{j},
    \label{eq:derivation-native-gate}
\end{aligned}
\end{equation}
where we are interested in the intrinsic operation of our resource,
\begin{equation}
    \begin{aligned}
        & G_I \coloneq \sum_{k=0}^{d-1} G_k \ket{0_X} \bra{k_Z} = \frac{1}{\sqrt{d}} \sum_{j,k=0}^{d-1} e^{i \theta_{kj}} \ket{j_Z} \bra{k_Z}.
    \end{aligned}
    \label{eq:native-operation}
\end{equation}
Unitarity of $G_I$ requires that
\begin{equation}
\begin{aligned}
    G_I G_I^\dagger = \sum_{k=0}^{d-1} G_k \ket{0_X} \bra{0_X} G_k^\dagger \overset{!}{=} I_d,
\end{aligned}
\end{equation}
so $\{ G_k \ket{0_X} \}_{k=0}^{d-1}$ being an orthonormal basis, implying that
\begin{equation}
   \bra{0_X} G_j^\dagger G_k \ket{0_X} = \frac{1}{d} \sum_{m} e^{i (\theta_{km} - \theta_{jm})} = \delta_{j,k}.
\end{equation}
This is exactly our previous condition $c_{j,k} = \delta_{j,k}$, so the two-qudit gate $G_E$ being able to create a maximally entangled two-qudit resource state $G_E \ket{+} \ket{+}$. Conversely, if $G_I$ is unitary, $\{G_I\ket{k_Z} \}_{k=0}^{d-1} = \{ G_k \ket{0_X} \}_{k=0}^{d-1}$ is an orthonormal basis and we have, thus, found a Schmidt decomposition \cite{NielsenChuang} of the two-qudit resource state,
\begin{equation}
    G_E \ket{+} \ket{+} = \frac{1}{\sqrt{d}} \sum_{k=0}^{d-1} \ket{k_Z} G_I \ket{k_Z} = (I \otimes G_I) \ket{\Phi},
    \label{eq:Schmidt-decomposition}
\end{equation}
where $\ket{\Phi} = \frac{1}{\sqrt{d}} \sum_k \ket{kk}$ is a Bell state.
Note that whenever the gate action of $G_E$ on the input is permutation symmetric, so $\theta_{jk} = \theta_{kj}$, we can shift the action of $G_E$ on the first qudit. This can also be seen from the more general Bell state property
\begin{equation}
    (I \otimes U) \ket{\Phi} = (U^T \otimes I) \ket{\Phi},
\end{equation}
for any unitary $U$, so if the transpose $(G_I)^T = G_I$, the intrinsic gate can be applied to either side.

In total, we have proven the equivalence of $G_E$ creating a maximally entangled two-qudit state $G_E \ket{+} \ket{+}$ and our resource state $G_E \ket{+} \ket{+}$ allowing for the Schmidt decomposition of Eq. \eqref{eq:Schmidt-decomposition} with a unitary $G_I$. Vice versa, a single-qudit unitary $G_I$ implies that $G_E \ket{+} \ket{+}$ is a maximally entangled two-qudit state.

From Eq. \eqref{eq:native-operation}, we can directly see that the choice $\theta_{jk} = \frac{2 \pi}{d} jk $, so using a controlled-phase gate for resource preparation, corresponds to an intrinsic Hadamard gate for the cluster state resource.

\subsubsection{Clifford intrinsic gate for Pauli by-product propagation}
\label{subsubsec:diagonal-entangling-Clifford-intrinsic-gate}

Given the two-qudit state $G_E \ket{\psi} \ket{+}$ with an arbitrary qudit input $\ket{\psi} = \sum_k \alpha_k \ket{k_Z}$, a measurement on the first qudit in the $X$ basis with outcome $j \in \mathbb{Z}_d$, realizes $G_I Z_d^{-j}$ since
\begin{equation}
\begin{aligned}
    & \bra{j_X}_1 \sum_{k=0}^{d-1} \alpha_k \ket{k_Z}_1 G_I \ket{k_Z}_2 
    \propto G_I Z^{-j} \sum_{k=0}^{d-1} \alpha_k \ket{k_Z}_2.
\end{aligned}
\end{equation}
If the intrinsic operation $G_I$ is Clifford, we can propagate an undesired Pauli by-product, originating from $Z^{-j}$ being introduced with each single-qudit measurement, to the beginning, adapting non-Clifford diagonal gates depending on previous measurement outcomes.

Hence, for universal measurement-based quantum computing, it is necessary to study which entangling gates $G_E$ are associated with a Clifford intrinsic gate $G_I$ of the resource state. We show that whenever the two-qudit diagonal entangling gate $G_E$ is Clifford, the intrinsic single-qudit gate $G_I$ is also a Clifford operation and vice versa. This will be achieved by deriving the stabilizer group generators of a two-qudit resource state in two different ways, where the first relies on the intrinsic gate and the second derivation uses the entangling gate.

The qudit Bell state $\ket{\Phi}$ has stabilizer group generators $X_d \otimes X_d$ and $Z_d \otimes Z_d^{-1}$ (for finite-field qudits, this becomes $X(x) \otimes X(x)$ and $Z(z) \otimes Z(-z)$ for all $x,z \in \mathbb{F}_d$) which are transformed by the intrinsic gate $G_I$ according to the Schmidt decomposition
\begin{equation}
    G_E \ket{+} \ket{+} = \frac{1}{\sqrt{d}} \sum_k \ket{k_Z} G_I \ket{k_Z}.
\end{equation}
Hence, if $G_I$ is Clifford, the resulting state is stabilized by the group, generated by $G_I X_d G_I^\dagger \otimes X_d$ and $G_I Z_d G_I^\dagger \otimes Z_d$.

In turn, if the entangling gate $G_E$ is a two-qudit Clifford operation, it modifies the stabilizer group generators of $\ket{+}^{\otimes 2}$, so $X_d \otimes I$ and $I \otimes X_d$, to $G_E(X_d \otimes I) G_E^\dagger$ and $G_E(I \otimes X_d) G_E^\dagger$. Since $G_E$ commutes with other diagonal gates including $Z_d$, for it to be Clifford only the conjugation of Pauli $X_d$ is relevant.

Thus, the stabilizer group of the two-qudit resource stabilizer state is given by
\begin{equation}
\begin{aligned}
    & \langle G_E(X_d \otimes I) G_E^\dagger, G_E(I \otimes X_d) G_E^\dagger \rangle
    \\ & = \langle G_I X_d G_I^\dagger \otimes X_d, G_I Z_d G_I^\dagger \otimes Z^{-1}_d \rangle,
\end{aligned}
\end{equation}
so that $G_E$ being Clifford is equivalent to $G_I$ being Clifford and vice versa.
Analogous arguments hold for $X(x)$ and $Z(z)$ in the prime-power-dimensional case.

\subsubsection{Universality of measurement-based gate set}
\label{subsec:universality-diagonal-entangling-gates}

We have to ensure that our generalized qudit stabilizer state resources allow the realization of a universal gate set. This particularly concerns the ability to perform arbitrary single-qudit unitaries via measurements on one-dimensional resource chains as in Fig. \ref{fig:setting} $(b)$.

The main idea of the proof is to expand an arbitrary single-qudit unitary into a Hermitian basis $\{ N_k \}_{k \in \mathbb{Z}_{d^2}}$
\begin{equation}
    U = e^{iH} = e^{i\sum_{k=1}^{d^2} \alpha_k N_k} = \prod_{k=1}^{d^2} e^{i \beta_k N_k}
    \label{eq:unitary-decomposition}
\end{equation}
for some real parameters $\alpha_k,\beta_k$ (the order of the product can be chosen as desired, however, this generally affects $\beta_k$) \cite{Clark_2006, Zhou_2003, LieAlgebraExponentialProducts} and show that all generators $\{ e^{i \beta N_k} \}_{k \in \mathbb{Z}_{d^2}}$ can be implemented measurement-based for all real $\beta$.

For finite-field qudits we explicitly show in Appendix \ref{app:universality-finite-field} with the help of previous results \cite{LieAlgebraExponentialProducts,appleby2009propertiesextendedcliffordgroup,MUBRef21,DURT_2010,Asadian_2016} how to decompose an arbitrary single-qudit unitary having the measurement-based gate set $\{ G_I, D_{\Vec{\phi}} \mid \phi \in \mathbb{R}^d \}$ available. Herein, the intrinsic gate $G_I$ is realized via single-qudit measurements in the $X$ basis, and $G_I D_{\Vec{\phi}}$ via single-qudit measurements in a rotated $X$ basis.
Then, $d ^2 - 1$ linearly independent Hermitian operators can be constructed from eigenstates of the $d+1$ Paulis $Z(1)$ and \cite{MUBRef21}
\begin{equation}
    \{ X(1) Z(a) \mid a \in \mathbb{F}_{d} \},
    \label{eq:disjoint-eigensets}
\end{equation}
yielding the Hermitian operators $\{ \ket{k_P} \bra{k_P} \}_{k \in \mathbb{Z}_{d-1}}$ for each Pauli $P$ in the above set, where $\ket{k_P}$ is an eigenstate of $P$ (we omit one of its $d$ eigenstates). Since the eigenstates of any Pauli operator $P$ are orthonormal, so the associated Hermitian operators commute, we can group $d-1$ generators according to
\begin{equation}
    \prod_k e^{i \alpha_k \ket{k_P} \bra{k_P}} =  e^{i \sum_k \alpha_k \ket{k_P} \bra{k_P}}.
    \label{eq:summarizing-generators}
\end{equation}
This simultaneous realization of multiple generators $\{ e^{i \beta N_k} \}_{k \in \mathbb{Z}_{d^2}}$ leads to a more efficient implementation of single-qudit gates.

\begin{figure}[]
    \centering
    \begin{tikzpicture}
    
    \coordinate (1) at (-1.8,0);  
    \coordinate (2) at (0,0);  
    \coordinate (3) at (1.8,0);  
    \coordinate (4) at (3.6,0);
    \coordinate (5) at (5.4,0);

    \draw[thick] (1) -- (5);

    \foreach \i in {1,2,3,4,5} {
        \node[circle, fill=pistacchio, draw=black] at (\i) {};
    }
    \node at (-0.9,0.3) {$G_E$};
    \node at (0.9,0.3) {$G_E$};
    \node at (2.7,0.3) {$G_E$};
    \node at (4.5,0.3) {$G_E$};
    
    \node at (-1.7,0.5) {$\ket{+}_1$};
    \node at (0.1,0.5) {$\ket{+}_2$};
    \node at (1.9,0.5) {$\ket{+}_3$};
    \node at (3.7,0.5) {$\ket{+}_4$};
    \node at (5.5,0.5) {$\ket{+}_5$};

    \node at (-1.7,-0.45) {$G_I S^\dagger D_{\alpha}$};
    \node at (0.1,-0.5) {$G_I D_{\beta} $};
    \node at (1.9,-0.5) {$G_I S $};
    \node at (3.7,-0.5) {$G_I D_{\gamma} $};
\end{tikzpicture}
    \caption{Implementation of a single-qubit unitary for generalized qubit resources, created by applying a diagonal Clifford $G_E$. If $G_I \propto G_I^{\dagger}$, so if the Pauli order of $G_I$ is two, four single-qubit measurements suffice. Below each qubit (green) in the figure, the gate executed with the respective single-qubit measurement is specified. For instance, measuring in the $X$ basis, rotated by $S^\dagger D_{\alpha}$ results in $G_I S^\dagger D_{\alpha}$ being realized. The fifth qubit, the output, then contains the state $ G_I D_{\gamma} G_I (S G_I D_{\beta} G_I S^\dagger) D_{\alpha} \ket{\psi}$, assuming zero outcome for all four measurements.}
    \label{fig:implementing-singlequditunitary}
\end{figure}
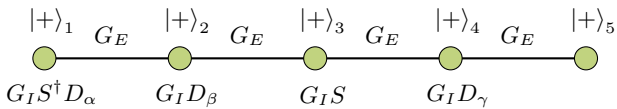

Next, we require that $G_I$ is a Clifford gate (in the sense of belonging to the unitary normaliser) that does not map the computational basis onto itself, so it conjugates $Z(z) \mapsto Z(a(z)) X(b(z))$ with $a, b \in \mathbb{F}_d$ and $b(z) \neq 0$ (up to a phase) for all $0 \neq z \in \mathbb{F}_d$.
This has two different reasons: first, having a Clifford intrinsic gate, we can propagate any Pauli by-product to the beginning of the computation and achieve determinism. Second, our universality proof assumes this specific Clifford operation is available.
For diagonal Clifford entangling gates, $b \neq 0$ is automatically satisfied. Explicitly, as derived in Appendix \ref{app:universality-finite-field}, one can then write any single-qudit unitary via
\begin{equation}
\begin{aligned}
    G_I D_{\Vec{\gamma}} G_I^\dagger \left( \prod_{\lambda \in \mathbb{F}_d \backslash \{ 0\}} S(\lambda) G_I D_{\Vec{\beta}(\lambda)} G_I^\dagger S(\lambda)^\dagger \right) D_{\Vec{\alpha}},
    \label{eq:unitary-decomposition-primepower}
\end{aligned}
\end{equation}
where $D_{\Vec{\alpha}} = e^{i \sum_k \alpha_k \ket{k_Z} \bra{k_Z}} $ with real parameters $\{ \alpha_k \}_k$ and analogous for the $d-1$ diagonal matrices $D_{\Vec{\beta}(\lambda)}$ and $D_{\Vec{\gamma}}$.
For odd prime-power dimensions, $S(\lambda)$ is $S_d^F(\lambda)$ from Eq. \eqref{eq:finitefield-phasegate-odd}, whereas for $p=2$, it is $M(\lambda) S_d^F M(\lambda^{-1})$ with $S_d^F$ from Eq. \eqref{eq:finitefield-phasegate-even}.
In Fig. \ref{fig:implementing-singlequditunitary}, we show how a single-qubit unitary would be realized in the simplest case, where the Pauli order of $G_I$ is two, via four measurements. Here, $o_{G_I}^P$ is the Pauli order of the intrinsic gate $G_I$, so the power that yields a Pauli operator up to a phase, allowing to obtain the inverse $G_I^\dagger$ via exponentiation of $G_I$.

Using the decomposition from Eq. \eqref{eq:unitary-decomposition-primepower}, we additionally derive an upper bound on the length of the measurement pattern required to realize any single-qudit unitary up to a Pauli by-product, namely $o_{G_I}^P d$. Thus, ideally, an optimal resource would have a self-inverse intrinsic gate, which is the case for the qudit cluster state for $p=2$. In general dimensions, it is, however, not possible to have an intrinsic gate being both self-inverse and satisfying the universality condition, as we discuss in Appendix \ref{app:self-inverse-intrinsic}.

For finite-dimensional integer-ring qudits, we demonstrate in Appendix \ref{app:universality-integer-ring} that the gate set $\{ G_I, D_{\Vec{\phi}} \mid \phi \in \mathbb{R}^d \}$ is exactly universal to realize arbitrary single-qudit unitaries if $G_I$ conjugates $Z_d \mapsto (Z_d)^a (X_d)^b$ with $b \in \mathbb{Z}_d^*$, so $b$ being invertible (which is equivalent to the greatest common divisor of $b$ and $d$ being $\gcd(b,d)=1$). This is achieved by showing that a Hadamard-like matrix can be constructed from the gate set $\{ G_I, S^k \mid k \in \mathbb{Z}_d \}$, so that all Clifford transformations can be realized \cite{Farinholt_2014}.

Working in the integer ring $\mathbb{Z}_d$, a Hermitian basis is no longer obtainable via projectors onto the eigenstates of the Pauli set in Eq. \eqref{eq:disjoint-eigensets}. Instead, each of the $d^2-1$ generators corresponds to $e^{i \sum_k \alpha_k \ket{k_P} \bra{k_P}}$ for appropriately chosen $d^2 -1$ Paulis $P$ and real parameters $\{\alpha_k\}_k$ \cite{Zhou_2003, Asadian_2016}. Here, $\ket{k_P}$ is an eigenstate of $P$. Moreover, arbitrary Clifford gates, required for the conjugation from $\ket{k_Z} \bra{k_Z}$ to $\ket{k_P} \bra{k_P}$, may now require sub-quadratic many Hadamard and phase gates (for prime dimensions, at most linearly many) \cite{Farinholt_2014}. Both the increased difficulty in the Hermitian basis and the decomposition of arbitrary Clifford gates make the decomposition of Eq. \eqref{eq:unitary-decomposition} more complicated than the one of Eq. \eqref{eq:unitary-decomposition-primepower} for finite-field qudits. Therefore, the number of required measurements to realize an arbitrary single-qudit unitary may be at the order of $d^4 o_{G_I}^P$ (more precisely, sub-quartic in $d$).

The derived decomposition of the Hadamard gate into the intrinsic gate $G_I$ and phase gates is also valid for finite-field qudits and tells us that Clifford circuits can be executed with simultaneous measurements without the need to adapt measurement bases since the Hadamard gate and phase gates generate (up to Paulis) the single-qudit Clifford group \cite{Farinholt_2014}. Note that even though the multiplication gate is often mentioned as a Clifford group generator both for integer-ring \cite{Farinholt_2014} and finite-field qudits \cite{heinrich2021stabiliser}, it can be expressed via Hadamard and phase gates \cite{Farinholt_2014}, as discussed in Appendix \ref{app:clifford-circuits}.

\begin{figure}[t]
    \centering
    \begin{tikzpicture}
    \foreach \x in {-2, -0.5, 1, 2.5, 4} {
        \foreach \y in {0,-1,-2} {
            \coordinate (\x\y) at (\x,\y);
        }
    }
      
    \foreach \x in {-2,-0.5, 1, 2.5} {
        \foreach \y in {0,-1,-2} {
            \draw [-stealth](\x , \y) -- (\x+1.5-0.15, \y); 
        }
    }

    \foreach \x in {-2,-0.5, 1, 2.5,4} {
        \foreach \y in {0,-1 } {
            \draw [-stealth](\x , \y) -- (\x, \y - 1 + 0.15); 
        }
    }

    \foreach \x in {-2,-0.5, 1, 2.5, 4} {
        \foreach \y in {0,-1,-2} {
            \node[circle, fill=pistacchio, draw=black] at (\x,\y) {}; 
        }
    }
\end{tikzpicture}
    \caption{Two-dimensional resource state for diagonal entangling interaction on resource qudits, initialized in $\ket{+}$. The horizontal lines allow for the implementation of single-qudit unitaries and the vertical lines allow for two-qudit entangling gates. The directions on the vertical lines can, in principle, be chosen freely since that only affects the local Clifford operations required for vertex deletion or local complementation. Here, we chose a symmetric configuration that allows the propagation of quantum information along vertical lines.
    }
    \label{fig:2D-resource-diagonal-gates}
\end{figure}
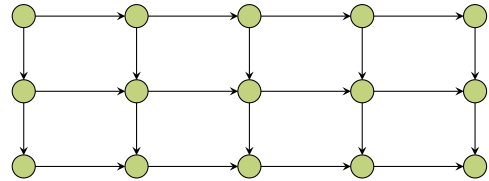

The missing piece for universal quantum computing is the implementation of a two-qudit entangling gate as in Fig. \ref{fig:setting} $(c)$ and single-qudit measurements that cut out required lattice structures from a universal resource state as in Fig. \ref{fig:setting} $(d)$. For this, we consider the two-dimensional lattice structure, shown in Fig. \ref{fig:2D-resource-diagonal-gates}, where the directed edges indicate which qudit acts as a control and which as a target of the corresponding diagonal entangling gate.
We discuss in Appendix \ref{app:LC-equivalent-cluster-state} that the resulting stabilizer state resource can be translated with local diagonal Clifford operations $C_1, C_2$ to a graph state with the same entanglement structure due to any diagonal Clifford entangling gate satisfying
\begin{equation}
    G_E = (C_1 \otimes C_2) (CZ)^N
    \label{eq:diagonal-entangling-clifford}
\end{equation}
for some integer $N \in \mathbb{Z}_d$ (for finite-field qudits, $M(N)CZM(N)$ with $N \in \mathbb{F}_d$, respectively), so that standard graph state manipulation tools are at our disposal \cite{hein2006entanglementgraphstatesapplications}. In particular, measurements in the $Z$ basis achieve vertex deletion together with the removal of all attached edges.
To create edges via a Pauli measurement on a vertex with $|N^o|$ outgoing and $|N^i|$ incoming edges, we can measure it in the $X Z^N$ basis (or the common eigenbasis of $X(x) Z(Nx)$ for all $x \in \mathbb{F}_d$ in case of finite-field qudits), conjugated by $(C_1)^{|N^o|} (C_2)^{|N^i|}$. More details, including the required measurement-dependent corrections on the neighborhood of the vertex, are discussed in Appendix \ref{app:resource-graph-manipulations}.

\subsubsection{Example: Light-shift gate resources}
\label{sec:lightshift-resources}

As an example, we now consider qudit resource states whose entanglement structure is characterized by the so-called light-shift gate, used to experimentally demonstrate an implementation of a native two-qudit entangling gate in a trapped-ion qudit system \cite{nativequditentanglinggate}. The light-shift gate $G(\theta)$ is defined by its action $G(\theta)\ket{jk} = e^{i \theta_{jk}} \ket{jk}$ with 
\begin{equation}
    \theta_{jk} = \theta (1- \delta_{j,k}) = \begin{cases}
    0 \quad, j = k\\
    \theta \quad, j \neq k
\end{cases}
\end{equation}
and applies a phase $e^{i\theta} $ for $j \neq k$ and acts as the identity otherwise.

The gates $G(\theta)$ commute due to being diagonal, so, as for controlled phase gates $CZ$, these resource states can be visualized without specifying a temporal order. Furthermore, due to the permutation symmetry on the input, $\theta_{jk} = \theta_{kj}$, one can use undirected graphs.

The reduced density matrix of $G(\theta) \ket{+} \ket{+}$ is
\begin{equation}
\rho = \frac{1}{d} \left(
    \begin{matrix}
        1 & c(\theta) & \hdots& c(\theta) \\ c(\theta) & \ddots & & \vdots \\ \vdots && \ddots & c(\theta) \\ c(\theta) & \hdots & c(\theta) & 1
    \end{matrix} \right),
\end{equation}
where according to Eq. \eqref{eq:off-diagonal-reduced-density-matrix}
\begin{equation}
    c(\theta) = \frac{1}{d}\left(e^{i\theta} + e^{-i\theta}+ d - 2 \right)= \frac{2 \cos(\theta) + d -2 }{d}.
\end{equation}
The prepared two-qudit resource state becomes maximally entangled if the reduced density matrix $\rho$ is maximally mixed, so if $c(\theta)=0$ or, equivalently, $\cos(\theta) = 1- d/2$. This condition can be satisfied for $d \in \{ 2,3,4\}$ since $\cos(\theta) \in [ -1,1 ]$. In particular, for $d=2$, we get $\theta = \pi/2$, for $d = 3$, we find that $\theta = 2\pi /3 $ produces a maximally entangled state so $e^{i\theta} = \omega_{3}$, and for $d = 4$, we obtain $\theta = \pi$.

For qubits, applying $G(\pi/2)$ onto $\ket{+}\ket{+}$, results in the intrinsic gate $U_2 \coloneq SHS \propto H S H$. As for the cluster state resource, measuring a qubit twice in the $X$ basis transports the state along a one-dimensional resource chain up to a Pauli by-product since the Pauli order of $U_2$ is two or, more explicitly,
\begin{equation}
    (U_2)^2 = (SHS)^2 = S H Z H S = X.
\end{equation}
The intrinsic gate $U_2$ leaves $X$ invariant and conjugates $U_2 Z U_2^\dagger = iXZ = Y $ so that all single-qubit gates are realizable according to our previously established criterion.

The qubit light-shift gate can be expressed via $G(\pi/2) = (S \otimes S) CZ$. The conjugation relations of $G(\pi/2)$ then allow us to obtain the stabilizer group generators of the associated resource stabilizer state. The description becomes slightly more complicated than for the cluster state since the stabilizer group generators now depend on the explicit structure of the graph, in particular, on the number of neighbors that each vertex has. Still, due to the local Clifford equivalence to a graph state (here, even a cluster state due to $N=1$ in Eq. \eqref{eq:diagonal-entangling-clifford}), vertex deletion is achieved via $Z$ measurements and edge creation via measurements in the $Y$ basis analogously to standard measurement-based quantum computing, as discussed in Appendix \ref{app:resource-graph-manipulations}, yielding a universal resource.

Applying the qutrit light-shift gate $G \left(\frac{2 \pi}{3} \right)$, which can be expressed via $(S_3^2 \otimes S_3^2) CZ_3$, results in the intrinsic Clifford gate
\begin{equation}
    U_3 \coloneq \frac{1}{\sqrt{3}} \left( \begin{array}{ccc}
         1& \omega_{3} & \omega_{3} \\
         \omega_{3}& 1& \omega_{3}\\
         \omega_{3} & \omega_{3} & 1
    \end{array} \right)  = i H_3 S_3 H_3^{-1} = S_3^2 H_3 S_3^2,
\end{equation}
which leaves $X_3$ invariant and conjugates $U_3 Z_3 U_3^\dagger = \omega_{3} X_3^{-1} Z_3$. Hence, the intrinsic gate allows for universal measurement-based quantum computing.

The Pauli order of $U_3$ is three due to $(U_3)^3 \propto I_3$ so that we can transport qutrits with three subsequent $X$ measurements. Note that a qutrit cluster state resource requires four measurements for transport due to $(H_3)^4 = I_3$. In particular, this also makes the decomposition of single-qutrit unitaries and, thus, quantum information processing, more efficient for the qutrit cluster state if it imitates the qutrit light-shift gate resource intrinsic gate $S_3^2 H_3 S_3^2$ via adjusting measurement bases for two subsequent measurements.

Applying the light-shift gate $G(\pi)$ on two ququarts creates a maximally entangled resource, characterized by the intrinsic gate
\begin{equation}
    U_4 \coloneq \frac{1}{2} \left( \begin{array}{cccc}
         1& -1& -1& -1 \\
         -1& 1& -1& -1 \\
         -1 & -1& 1& -1 \\
         -1& -1 & -1& 1
    \end{array} \right) = H_4^F \left( \begin{array}{cccc}
         -1 &  \\
         & 1 \\
         & & 1 \\
         & & & 1
    \end{array} \right) H_4^F,
    \label{eq:U4-decomposition}
\end{equation}
where $(U_4)^2 = I_4$. Hence, the Pauli order is minimal as for the ququart finite-field cluster state. In particular, while $X(x)$ remains unaffected, the finite-field Pauli $Z(z)$ gates are conjugated according to
\begin{equation}
    U_4 Z(z) U_4 = - Z(z) X(z^{-1})
    \label{eq:U4-Zconjugation}
\end{equation}
for all $x, z \in \mathbb{F}_4$, $z \neq 0$, so that, with our previously derived criterion, $U_4$ is able to support a universal gate set.

The ququart light-shift gate maps the Heisenberg-Weyl group onto itself, however, due to $z \mapsto z^{-1}$ in Eq. \eqref{eq:U4-Zconjugation}, the finite-field argument does not transform linearly. Hence, we cannot express $G(\pi)$ via two-ququart Clifford gates but only via four-qubit Clifford operations according to
\begin{equation}
    G(\pi) = Z^{\otimes 4} CZ_{12} CZ_{23} CZ_{34} CZ_{14}.
\end{equation}
Here, we use the natural embedding of two ququarts into four qubits via
\begin{equation}
    \ket{a_2 + a_1 \xi } \ket{a_4 + a_3 \xi} \mapsto \ket{a_1} \ket{a_2} \ket{a_3} \ket{a_4},
\end{equation}
where $a_1, a_2, a_3, a_4 \in \mathbb{Z}_2$.

Similarly, the intrinsic gate $U_4$ cannot be expressed via $D_2 H_4^F M(w) D_1$ for any $w \in \mathbb{F}_4 $ and any diagonal unitaries $ D_1, D_2$, so that it cannot be understood as a finite-field ququart graph state with adjusted measurement bases. Considering the two-qubit Clifford group (the full unitary normalizer of a single ququart), yields
\begin{equation}
    U_4 = (H \otimes H) (X \otimes X) CZ (X \otimes X) (H \otimes H).
    \label{eq:lightshift-ququart-intrinsic}
\end{equation}

In principle, ququarts can be described both in terms of the integer ring $\mathbb{Z}_4$ and the finite field $\mathbb{F}_4$. However, choosing an inter-ring perspective and computing conjugation relations, we find that the intrinsic operation $U_4$ is not an integer-ring Clifford gate.

\subsection{Block-diagonal Clifford entangling gates}
\label{sec:blockdiagonal-entangling-gate}

\subsubsection{Intrinsic gate associated with resource}

We now consider the class of block-diagonal entangling gates
\begin{equation}
    G_E \coloneq \sum_k \ket{k} \bra{k} \otimes U_k,
    \label{eq:block-diagonal}
\end{equation}
where the previous class of diagonal gates is included via $U_k = G_k$ being diagonal. We consider this resource class since it admits the measurement-based implementation of an entangling gate via mediator qudits, as discussed later.

\begin{figure}[t]
    \centering
    \begin{tikzpicture}

    \coordinate (1) at (-2,0); 
    \coordinate (2) at (0,0);  
    \coordinate (3) at (2,0); 
    \coordinate (4) at (4,0);

    \draw [-stealth](1) -- (-0.15,0);
    \draw [-stealth](2) -- (2-0.15,0);
    \draw [-stealth](3) -- (4-0.15,0);

    \foreach \i in {1,2,3,4} {
        \node[circle, fill=pistacchio, draw=black] at (\i) {};
    }
    
    \node at (-0.8,-0.4) {$1)$ $ G_E$};
    \node at (1.2,-0.4) {$2)$ $ G_E$};
    \node at (3.2,-0.4) {$3)$ $ G_E$};

    \node at (-1.9,0.5) {$ \ket{\varphi}$};
    \node at (0.1,0.5) {$ \ket{\varphi}$};
    \node at (4.1,0.5) {$ \ket{\varphi}$};
    \node at (2.1,0.5) {$ \ket{\varphi}$};
\end{tikzpicture}
    \caption{One-dimensional resource chain for single-qudit unitary implementation. The resource qudits are initialized in $\ket{\varphi} = D_{\Vec{\varphi}} \ket{+}$ and block-diagonal entangling gates $G_E$ applied from left to right, using the left qudit as the control and the right qudit as the target.
    }
    \label{fig:onedimensional-resource}
\end{figure}
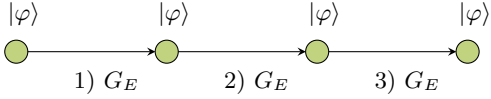

To implement single-qudit gates, we make use of one-dimensional chains, Fig. \ref{fig:onedimensional-resource}. Due to the entangling gates no longer commuting in general (they only commute on the same control qudit), we need to specify a temporal order for their application.

If we now apply $G_E$ on $\ket{\psi} \ket{\varphi}$ with a resource qudit initialized in $\ket{\varphi}$ and an input state $\ket{\psi} = \sum_j \alpha_j  \ket{j}$, we obtain by definition
\begin{equation}
\begin{aligned}
     & \sum_k  \bra{k} \ket{\psi} \ket{k} U_k \ket{\varphi} = \sum_{k} \alpha_k \ket{k} \left( \sum_j U_j \ket{\varphi} \bra{j} \right) \ket{k},
\end{aligned}
\end{equation}
so that the intrinsic operation of the resource state becomes
\begin{equation}
    G_I \coloneq \sum_j U_j \ket{\varphi} \bra{j}.
\end{equation}
Since we want to process the qudit input with a unitary, we demand that 
\begin{equation}
\begin{aligned}
    & I_d = G_I G_I^\dagger = \sum_{j,k} U_j \ket{\varphi} \bra{j} \ket{k} \bra{\varphi} U_k^\dagger
     = \sum_{k} U_k \ket{\varphi} \bra{\varphi} U_k^\dagger,
\end{aligned} 
\end{equation}
so that $\{ U_k \ket{\varphi} \}_k$ is an orthonormal basis.

Applying a block-diagonal entangling gate on $\ket{\varphi} \ket{\varphi}$ yields the two-qudit resource state
\begin{equation}
\begin{aligned}
     & \sum_k  \bra{k} \ket{\varphi} \ket{k} U_k \ket{\varphi} = \sum_{k} \bra{k} \ket{\varphi} \ket{k} G_I \ket{k}.
\end{aligned}
\end{equation}
If $G_I$ is unitary this is the Schmidt decomposition \cite{NielsenChuang} since we can always write $\ket{\varphi} = \sum_k \alpha_k e^{i \varphi_k} \ket{k_Z}$ with real $\alpha_k$ satisfying $\sum_k \alpha_k^2 =1$ due to normalization, so that the parameters $\{ \alpha_k \}_k$ exactly correspond to the Schmidt coefficients.

The two-qudit resource is then a maximally entangled state in case of equal Schmidt coefficients, so $\alpha_k = \frac{1}{\sqrt{d}}$ for all $k$. A trivial possibility for this is $\ket{\varphi} = \ket{0_X}$. However, also if $\ket{\varphi} = D_{\Vec{\varphi}} \ket{0_X}$ with a diagonal unitary $D_{\Vec{\varphi}}$, we can write
\begin{equation}
\begin{aligned}
     & \sum_k  \bra{k} D_{\Vec{\varphi}} \ket{0_X} \ket{k} U_k \ket{\varphi} = \sum_{k} \bra{k} \ket{0_X} D_{\Vec{\varphi}} \ket{k} G_I \ket{k}
     \\ & = \frac{1}{\sqrt{d}} \sum_{k} D_{\Vec{\varphi}} \ket{k} G_I \ket{k} = \frac{1}{\sqrt{d}} \sum_{k} \ket{k} G_I D_{\Vec{\varphi}} \ket{k}.
\end{aligned}
\label{eq:Schmidt-decomposition-block-diagonal}
\end{equation}
Hence, $\ket{\varphi} = D_{\Vec{\varphi}} \ket{0_X}$ with a diagonal $D_{\Vec{\varphi}}$ and $G_I$ unitary yields a maximally entangled state $G_E \ket{\varphi} \ket{\varphi}$.

\subsubsection{Clifford intrinsic gate and Pauli by-product propagation}

As previously, we would like our intrinsic gate to be Clifford. This will allow us to rely on the same universality proof for decomposing arbitrary single-qudit unitaries as in Sec. \ref{subsec:universality-diagonal-entangling-gates} and propagate Pauli by-products to the beginning of the computation.
Analogous to Sec. \ref{subsubsec:diagonal-entangling-Clifford-intrinsic-gate}, we derive the stabilizer group generators of a two-qudit resource state in two different ways to understand the Clifford properties of the entangling and the intrinsic gates.

We start with a two-qudit resource $G_E \ket{\varphi} \ket{\varphi}$, where $\ket{\varphi} = D_{\Vec{\varphi}} \ket{0_X}$, and a block-diagonal entangling gate $G_E$ as in Eq. \eqref{eq:block-diagonal}. The purpose of introducing $D_{\Vec{\phi}}$ lies in block-diagonal entangling gates not always producing an entangled state when applied onto $\ket{++}$ (for instance, $CX \ket{++}$ is not entangled). However, we require an entangled two-qudit resource state for unitarity of $G_I$, such that preparing $\ket{\varphi} = D_{\Vec{\varphi}} \ket{+}$ instead gives us more flexibility in achieving an intrinsic gate that allows for universal measurement-based quantum computing. Notice that the only diagonal single-qudit Clifford gate corresponds to $S$ or powers thereof since a diagonal gate always commutes with $Z$ and can only map $X \mapsto X Z^l$ for some $l$, which is already achieved by $S^l$.

In the Schmidt decomposition of Eq. \eqref{eq:Schmidt-decomposition-block-diagonal}, we observe that in the case of $G_I$ and $D_{\Vec{\varphi}}$ being Clifford, the stabilizer group of the two-qudit stabilizer resource state becomes
\begin{equation}
    \langle D_{\Vec{\varphi}} X_d D_{\Vec{\varphi}}^\dagger \otimes G_I  X_d  G_I^\dagger, Z_d^{-1} \otimes G_I Z_d G_I^\dagger \rangle
\end{equation}
due to the conjugation of the Bell state stabilizers $X_d \otimes X_d$ and $Z_d^{-1} \otimes Z_d$ with $D_{\Vec{\varphi}} \otimes G_I$.

On the other hand, applying the entangling gate to the resource $\ket{\varphi}^{\otimes 2}=(D_{\Vec{\varphi}} \ket{0_X})^{\otimes 2}$ leads to the same state, so the same stabilizer group. Before applying $G_E$, the stabilizer group generators of $\ket{\varphi} \ket{\varphi}$ are $D_{\Vec{\varphi}} X_d D_{\Vec{\varphi}}^\dagger \otimes I$ and $I \otimes D_{\Vec{\varphi}} X_d D_{\Vec{\varphi}}^\dagger$, so that afterwards we obtain the stabilizer group
\begin{equation}
\begin{aligned}
        & \langle G_E(D_{\Vec{\varphi}} X D_{\Vec{\varphi}}^\dagger \otimes I)G_E^\dagger, G_E(I \otimes D_{\Vec{\varphi}} X D_{\Vec{\varphi}}^\dagger) G_E^\dagger \rangle.
\end{aligned}
\end{equation}

Hence, if $G_E(D_{\Vec{\varphi}} \otimes I)$ and $G_E(I \otimes D_{\Vec{\varphi}})$ are Clifford (less restricted, they conjugate $X_d \otimes I$ and $I \otimes X_d$ to a Pauli), also $G_I $ and $D_{\Vec{\varphi}}$ are Clifford.

\subsubsection{Universality of resource state}

\begin{figure}[t]
    \centering
    \begin{tikzpicture}

    \foreach \x in {-2, -0.5, 1, 2.5, 4} {
        \foreach \y in {0,-1,-2} {
            \coordinate (\x\y) at (\x,\y);
        }
    }
      
    \foreach \x in {-2,-0.5, 1, 2.5} {
        \foreach \y in {0,-2} {
            \draw [-stealth](\x , \y) -- (\x+1.5-0.15, \y); 
        }
    }

    \foreach \x in {-2,-0.5, 1, 2.5,4} {
        \draw [-stealth](\x , 0) -- (\x,  -1 + 0.15);
        \draw [-stealth](\x , -2) -- (\x,  -1 - 0.15); 
    }

    \foreach \x in {-2,-0.5, 1, 2.5, 4} {
        \foreach \y in {0,-2} {
            \node[circle, fill=pistacchio, draw=black] at (\x,\y) {}; 
        }
    }

    \foreach \x in {-2,-0.5, 1, 2.5, 4} {
        \node[circle, fill=mutedplum, draw=black] at (\x,-1) {}; 
    }
\end{tikzpicture}
    \caption{Two-dimensional resource state for block-diagonal Clifford entangling gates, allowing the implementation of arbitrary single-qudit unitaries along horizontal chains and the two-qudit entangling gate $(S \otimes S) CZ$ via the vertical connection. The horizontal one-dimensional chains are used for quantum information processing with single-qudit gates while single-qudit measurements on the mediator qudits (dark violet) in appropriate bases allow for disentangling or entangling two computational qudits (light green), as shown in Fig. \ref{fig:mediator-qudit}.
    }
    \label{fig:2D-resource-blockdiagonal}
\end{figure}
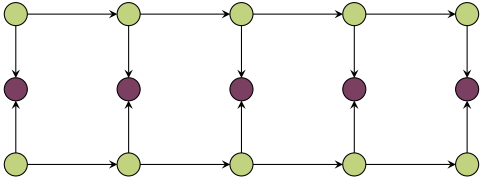

For Clifford intrinsic gates $G_I$ that conjugate the $Z$ basis to an invertible power of $X$, we can decompose arbitrary single-qudit unitaries according to Eq. \eqref{eq:unitary-decomposition}, as discussed in Sec. \ref{subsec:universality-diagonal-entangling-gates}.
In addition to single-qudit unitaries, we require the ability to perform any two-qudit entangling gate for exact universal measurement-based quantum computing and to cut out required resource state structures, as in Fig. \ref{fig:setting} $(d)$.

For this, we consider the configuration in Fig. \ref{fig:2D-resource-blockdiagonal}.
In such a universal resource state, the horizontal lines allow the implementation of single-qudit unitaries on one-dimensional chains, as displayed in Fig. \ref{fig:setting} $(b)$. The subsequent vertical entangling interaction uses the computational qudits of horizontal lines as controls and mediator qudits in between horizontal lines as targets. Single-qudit measurements on the mediator qudits then allow either disconnecting neighboring horizontal lines or performing a two-qudit entangling gate between two computational qudits.

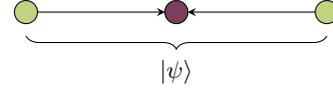
\begin{figure}[t]
    \centering
    \begin{tikzpicture}
    \coordinate (1) at (-2,0);  
    \coordinate (2) at (0,0);  
    \coordinate (3) at (2,0); 

    \draw [-stealth](1) -- (-0.15,0);
    \draw [-stealth](3) -- (0.15,0);

    \foreach \i in {1,3} {
        \node[circle, fill=pistacchio, draw=black] at (\i) {};
    }
    \node[circle, fill=mutedplum, draw=black] at (2) {};

    \draw [decorate,decoration={brace,amplitude=5pt,mirror,raise=9pt}]
    (1) -- (3) node [black,midway,yshift = -0.8cm] 
    {$\ket{\psi}$};
    
\end{tikzpicture}
    \caption{The intermediate qudit (dark violet) is used either as a mediator for an entangling gate acting on the two-qudit input $\ket{\psi}$ or to disconnect the computational qudits (light green) via measurements in appropriate bases.
    }
    \label{fig:mediator-qudit}
\end{figure}

To understand this, we consider the configuration in Fig. \ref{fig:mediator-qudit}, where an arbitrary two-qudit state $\sum_{j,k} \alpha_{j,k} \ket{j} \ket{k}$ has been transported and processed up to a vertical edge of Fig. \ref{fig:2D-resource-blockdiagonal}. As discussed in Appendix \ref{app:2Dresource-blockdiagonal}, we can express any block-diagonal Clifford entangling gate via
\begin{equation}
    (C_1 \otimes C_2) CP,
    \label{eq:structure-blockdiagonal}
\end{equation}
where $CP$ is a controlled-Pauli, $C_1$ a diagonal local Clifford gate, and $C_2$ an arbitrary local Clifford gate.

If both local Clifford gates in Eq. \eqref{eq:structure-blockdiagonal} are trivial, so $C_1 = C_2 = I_d$, and $P \propto Z^a X^b$ is a Pauli with the greatest common divisor $\gcd(a,b)$ being invertible, we can prepare the mediator qudit in an appropriate basis (explicitly derived in Appendix \ref{app:2Dresource-blockdiagonal}), such that applying a controlled-Pauli $CP$ results in the total state
\begin{equation}
    \sum_{k,j} \alpha_{k,j} \ket{k} \bra{k} \otimes G_C \ket{j+k} \otimes \ket{j} \bra{j},
\end{equation}
where $G_C$ is some Clifford gate. Then, a measurement of the mediator qudit in the basis $\{ G_C \ket{k_X} \}_k$ disentangles both computational qudits while a measurement in the basis $\{ G_C S^{-1} \ket{k_X} \}_k$ entangles both qudits via $(S \otimes S) CZ$ (leaving a measurement outcome-dependent by-product $Z^{-k} \otimes Z^{-k}$ in both cases).

A non-trivial local diagonal Clifford $C_1$ in the block-diagonal entangling gate of Eq. \eqref{eq:structure-blockdiagonal} allows for the same strategy. However, a non-trivial $C_2$ gate that does not commute with the Pauli $P$ disturbs the commutation of two entangling gates on the same target. Hence, in cases where the qudit stabilizer state resource is explicitly prepared (instead of arising via a natural interaction), the $C_2$ gate needs to be removed after the application of the first entangling gate to the mediator qudit. Otherwise, the previous decoupling strategy does not work. Another approach to potentially deal with the presence of a non-trivial $C_2$ gate that does not commute with $P$ in Eq. \eqref{eq:structure-blockdiagonal} is applying the entangling gate multiple times until $C_2$ vanishes and a controlled-$P$ gate with appropriate $P$ (so the greatest common divisor of the $X$ and $Z$ powers being invertible) is obtained. For instance, in case of an entangling qubit interaction $(I \otimes H) CZ$, applying the gate twice on the same control and target results in the entangling gate $CX \cdot CZ$, for which we can use the previous controlled-Pauli strategy.

Then, the structure in Fig. \ref{fig:2D-resource-blockdiagonal} allows the cut out of one-dimensional chains for single-qudit gate implementation as well as the implementation of the Clifford entangling gate $(S \otimes S) CZ$, similar to Fig. \ref{fig:setting} $(d)$.

\subsubsection{Allowing for an unknown quantum input}

In general, our stabilizer state resources have as an input the state $(D_{\Vec{\varphi}} \ket{0_X})^{\otimes n}$, which can be transformed into any other state for the computation being performed. However, it may be of interest to process an unknown qudit quantum state, which requires coupling it into a universal resource state. While this is simple for diagonal entangling gates, where one can simply apply the entangling gate that characterizes the resource state to the unknown qudit input and a qudit of the resource state, block-diagonal gates no longer commute and require a different approach.

Therefore, we consider a three-qudit scenario, where a qudit input $\ket{\psi}$ is connected to a two-qudit resource. We then have the total state $\ket{\psi}_1 (G_E)_{23} \ket{\varphi}_2 \ket{\varphi}_3$ with $\ket{\varphi} = D_{\Vec{\varphi}} \ket{0_X}$. Using the Schmidt decomposition of Eq. \eqref{eq:Schmidt-decomposition-block-diagonal}, this corresponds to
\begin{equation}
\begin{aligned}
    & \ket{\psi}_1 (G_E)_{23} \ket{\varphi}_2 \ket{\varphi}_3
    = \ket{\psi}_1 \frac{1}{\sqrt{d}} \sum_{j} D_{\Vec{\varphi}} \ket{j}_2 G_I \ket{j}_3.
\end{aligned}
\end{equation}
Now we remove $D_{\Vec{\varphi}}$ and subsequently perform a Bell state measurement on qudits one and two or, equivalently, measure in the Bell basis, rotated by $D_{\Vec{\varphi}}$ on qudit two. Here, the Bell state measurement has the possible outcomes \cite{Clark_2006}
\begin{equation}
    \{ \ket{\Phi(s,t)} = ( I \otimes X^s Z^t) \ket{\Phi} \mid s,t \in \mathbb{Z}_d \}.
\end{equation}
For the outcome $\ket{\Phi}_{12}$ and after re-normalization, we obtain $G_I \ket{\psi}_3$, while for the other possible outcomes, we additionally obtain a Pauli by-product $X^{k_1} Z^{k_2}$ that can be propagated.
Since the Bell state measurement is performed on qudits one and two, it commutes with entangling gates and measurements on the third qudit, so we have found a procedure to couple quantum information into one-dimensional resource state chains of arbitrary length.

\subsubsection{Example: Controlled-$X$ gate resources}
\label{subsec:CX-gate-preparation}

For resources, characterized by controlled-$X$ gates, the entangling gates do not commute unless they act on the same control or target. Thus, one needs to keep track of the temporal order in which the $CX$ gates are applied in addition to specifying the control and target of each operation.

We consider a one-dimensional resource state as shown in Fig. \ref{fig:onedimensional-resource}, initializing the resource qudits in $ \ket{\varphi} = S_d \ket{0_X}$. Applying afterwards the $CX_d$ gate, we obtain for integer-ring qudits
\begin{equation}
\begin{aligned}
     & \sum_k \alpha_k \ket{k} X_d^k S_d \ket{0_X}
      = \sum_k \alpha_k \ket{k} S_d \tau_d^{-k^2} X_d^{k} Z_d^{-k} \ket{0_X}
\end{aligned}
\end{equation}
by conjugation with $S_d^\dagger$.
Note that while $S_d X_d S_d^\dagger = \tau_d X_d Z_d$, we have that
\begin{equation}
    S_d X^k S_d^\dagger = \tau_d^k (X_d Z_d)^k = \tau_d^k \tau_d^{(k-1)k} X_d^k Z_d^k = \tau_d^{k^2} X_d^k Z_d^k
\end{equation}
due to $Z_d X_d = \omega_d X_d Z_d $. We then need to commute the second $X_d$ gate once and the last $k-1$ times, introducing each time $\omega_d = \tau_d^2$, so $\sum_{j=1}^{k-1} j = (k-1)k/2$ many times in total for $k \geq 2$. For the next conjugation, we observe that $\ket{0_X} = H_d \ket{0_Z} = H_d^{-1} \ket{0_Z}$, so that we can write
\begin{equation}
\begin{aligned}
    & \sum_k \alpha_k \ket{k} S_d H_d^{-1} \tau_d^{-k^2} Z_d^{k} X_d^{k} \ket{0}
     \\ & = \sum_k \alpha_k \ket{k} S_d H_d^{-1} \tau_d^{-k^2+2k^2} X_d^{k} Z_d^{k} \ket{0}
      \\ & = \sum_k \alpha_k \ket{k} S_d H_d^{-1} S_d \ket{k},
\end{aligned}
\end{equation}
obtaining the intrinsic gate $G_I = S_d H_d^{-1} S_d$.

For finite-field qudits with $p=2$, initialized in $\ket{\varphi} = S_d^F \ket{0_X}$, conjugation of $X(x)$ with $H_d^F (S_d^F)^\dagger$ yields $\chi_4(x^2) X(x) Z(x) $, so that the intrinsic operation becomes $S H^{-1} S $. Similarly, for $p\neq 2$ and resource qudit initialization in $\ket{\varphi} = S_d^F \ket{0_X}$, conjugation of $X(x)$ transforms it into $\chi(2^{-1} x^2 ) X(x) Z(x)$, resulting in the intrinsic gate $S(1) H^{-1} S(1)$.

Hence, the intrinsic operation is $SH^{-1}S$ irrespective of whether we consider integer-ring or finite-field qudits. Interestingly, for qubits, this equals $SHS$, which also appeared as the intrinsic operation of a qubit light-shift gate resource. So, while fixing an entangling gate and a qudit initialization uniquely characterizes the intrinsic gate associated with the resource, the reverse is not true.

Due to $SH^{-1} S$ conjugating $Z_d$ (or $Z(1)$, respectively) to $X_d Z_d$ (or $X(1) Z(1)$) up to a phase, having $S H^{-1} S$ as an intrinsic gate, arbitrary single-qudit unitaries can be implemented both for integer-ring and finite-field qudits.

Applying two $CX$ gates on an arbitrary two-qudit input $\sum_{l,j} \alpha_{l,j} \ket{l} \ket{j}$, where each qudit is used as a control, and on a mediator qudit initialized in $\ket{\phi} = S \ket{0_X}$ as a target of both gates, we can due to $G_I = S H^{-1} S$, express the action of the two $CX$-gates via
\begin{equation}
    \begin{aligned}
        \sum_{l,j} \alpha_{l,j} \ket{l} S H^{-1} S \ket{l+j} \ket{j}.
    \end{aligned}
\end{equation}
We now immediately see that a measurement in the $ \{S H^{-1} S \ket{k_X} \}_k$ basis on the mediator qudit disentangles the two computational qudits, leaving each with a $Z^{-k}$ by-product for the outcome $k \in \mathbb{Z}_d$. Hence, we can isolate horizontal lines and enable single-qudit gates in a measurement-based way.

In turn, a measurement in the basis $ \{ S H^{-1} \ket{k_X} \}_k = \{ S \ket{k_Z} \}_k$ leaves $\bra{0_X} S \ket{l+j}$, where
\begin{equation}
    S \ket{l+j} = \tau_d^{(l+j)^2} \ket{l+j} = \omega_d^{lj} \tau_d^{j^2+l^2} \ket{l+j}.
\end{equation}
Hence, we can rewrite the interaction on the computational qudits via
\begin{equation}
    CZ (S Z^{-k} \otimes S Z^{-k}) \sum_{l,j} \alpha_{l,j} \ket{l} \otimes \ket{j}.
\end{equation}
Thus, we are able to apply the Clifford entangling gate $CZ(S \otimes S)$ measurement-based and have a universal resource stabilizer state. In Appendix \ref{app:2Dresource-blockdiagonal}, we describe a similar strategy to realize $CZ(S \otimes S)$ for other block-diagonal Clifford entangling gates.

Moreover, since the intrinsic gate $S H^{-1} S$ leaves $X$ invariant, using the same initialization $\ket{\varphi} = S \ket{+}$ for an entangling interaction $(CX)^l$ simply changes the intrinsic gate to $S H^{-1} S M(l)$.

Interestingly, the $CX$-resource state example is equivalent to a qudit graph state via local diagonal Cliffords with the same entanglement connectivity and the edge weight $w = -1$ everywhere, which we demonstrate in Appendix \ref{app:CX-weight-1-graph} via a stabilizer group generator argument. This also explains $H^{-1}$ appearing in the intrinsic gate of the $CX$-resources since a qudit graph state with equal edge weight $w$ has the intrinsic gate $H M(w)$ and $H M(-1) = H^3 = H^{-1}$ in all dimensions. We discuss in the next section in which cases a qudit graph state can imitate a stabilizer state resource via adjusting measurement bases.

\section{Comparison of resources}
\label{sec:resource-comparison}

\subsection{Graph state-like resource states}

\subsubsection{Modifying the intrinsic gate}
\label{subsec:adapting-intrinsic-gate}

For a stabilizer state resource being locally and diagonally Clifford equivalent to a qudit graph state with the same entanglement connectivity and equal edge weight $w$, the graph state can mimic the respective stabilizer state resource by adjusting the measurement basis for two subsequent rotated $X$ measurements. In particular, the intrinsic gate of the qudit graph state can be effectively modified to $D_2 H M(w) D_1$ with $D_1$ and $D_2$ being diagonal gates. For $w=1$, the graph state resource then corresponds to the standard qudit cluster state.

We first consider the case that $D_1$ and $D_2$ are local diagonal Clifford gates, so that the modified intrinsic gate remains Clifford. Since the only diagonal Clifford group generators are $S(\lambda)$, $\lambda \in \mathbb{F}_d$, (or $S^{\lambda}$ with $\lambda \in \mathbb{Z}_d$ for integer-ring qudits) and they already produce all possible transformations of $X(x)$, the modified intrinsic gate of an equal-weight qudit graph state can be written via $S(s_2)H M(w) S(s_1)$. Hence, if $S(s_2)H M(w) S(s_1)$ has a lower Pauli order than $H M(w)$, we can reduce the measurement depth for realizing arbitrary single-qudit unitaries via the decomposition of Eq. \eqref{eq:unitary-decomposition-primepower} along one-dimensional resource state chains. This is the case for two qutrit resource state examples that we revisit in the next section.

Note that $S(s_2) H M(w) S(s_1)$ transforms the Pauli gates according to (ignoring phases)
\begin{equation}
\begin{aligned}
    & Z(z) \mapsto Z(-s_2 w^{-1} z) X(-w^{-1} z) 
    \\ & X(x) \mapsto Z(-s_1 s_2 w^{-1} x + wx) X(-s_1 w^{-1} x).
\end{aligned}
\end{equation}
Thus, if we have a finite-field Clifford intrinsic gate $G_I$ that transforms 
\begin{equation}
\begin{aligned}
    & Z(z) \mapsto Z(az) X(bz)
    \\ & X(x) \mapsto Z(cx) X(dx)
    \label{eq:clifford-primepower}
\end{aligned}
\end{equation}
with $b$ being invertible, such that we can do universal measurement-based quantum computing, we can express $c = (ad - 1)b^{-1}$ due to the condition $ad - bc =1$ that preserves Pauli commutation relations. Setting
\begin{equation}
    w \coloneq - b^{-1}, s_1 \coloneq d b^{-1}, s_2 \coloneq ab^{-1},
\end{equation}
a qudit graph state with weight $w$ for all edges and adjusted measurement bases via $S(s_1)$ for the first measurement and $S(s_2)$ for the second can then induce the same Pauli group transformation. Thus, all Clifford intrinsic gates, which can result in universal measurement-based quantum computing, are already exhausted by equal-weight qudit graph states. In turn, given a qudit graph state, one can optimize $s_1, s_2, w \in \mathbb{F}_d$ to achieve minimal Pauli order.

The above argument is only valid for Clifford gates that transform the Pauli argument linearly as in Eq. \eqref{eq:clifford-primepower}, whereas for universal measurement quantum computing, gates from the unitary normalizer of the Pauli group suffice in principle. For integer-ring qudits, the full normalizer of the Pauli group is already described by Clifford gates since the transformation of $Z$ and $X$ specifies the mapping of any $Z^a X^b$. Instead, finite-field qudits of prime-power dimension with non-unit exponent can have normalizer gates which preserve the Pauli group but with a non-linear transformation of the finite-field arguments $x,z \in \mathbb{F}_d$ of $Z(z)$ or $X(x)$. In this case, the resulting resource state can in general not be mimicked by a qudit graph state, even when permitting $D_1$ and $D_2$ to be normalizer gates, as we show in Appendix \ref{app:self-inverse-intrinsic}. This resource state class will be discussed in Sec. \ref{subsec:non-graph-state-like}, while next we compare resource state examples that can be mimicked by qudit graph states.

\subsubsection{Resource state examples}

\begin{table*}[t!]
\renewcommand{\arraystretch}{1.35}
\begin{tabular}{c|c|c|c|c}
Dimension $d$ & Entangling interaction & Intrinsic gate $G_I$ & Pauli order $o_{G_I}^P$ & Total cost \\ \hline \hline
Qubits $(d=2)$ & Controlled-$Z$ &  $H$ & 2 & $\leq 4$     \\
  & Light-shift & $S H S$ & 2 & $\leq 4$    \\
  & Controlled-$X$ & $S H S$   &  2 & $\leq 4$     \\ \hline
Qutrits $(d=3)$ & Controlled-$Z$ &  $H_3$ & 4 & $\leq 12$      \\
  & Light-shift & $ i H_3 S_3 H_3^{-1}$ $= S_3^2 H_3 S_3^2$ & 3 & $\leq 9$    \\
  & Controlled-$X$ & $S_3 H_3^{-1} S_3$   &  3 & $\leq 9$  \\ \hline
Ququarts $(d=4)$ & Controlled-$Z$ &  $H_4^F$ & 2 & $\leq 8$ \\
  & Light-shift & $H_4^F \textnormal{diag}(-1,1,1,1) H_4^F $ & 2 & $ \leq 8$  \\
  & Controlled-$X$ & $S_4^F (H_4^F)^{-1} S_4^F$   & 2 & $\leq 8$ \\ \hline
  Odd prime-power $p^m$ & Controlled-$Z$ &  $H_d^F$ & 4 & $\leq 4d$ \\
  & Controlled-$X$ & $S_d^F(1) (H_d^F)^{-1} S_d^F(1)$   & $p$ & $\leq p d$   \\
  & Block-diagonal Clifford & $-$   & $ \leq \max\{4p,d+1\}$  & $\leq d \max\{4p,d+1\} $ 
  \\ \hline
  Even prime-power $2^m$ & Controlled-$Z$ &  $H_d^F$ & 2 & $\leq 2d$    \\
  & Controlled-$X$ & $S_d^F H_d^F S_d^F$   &  2 & $\leq 2d$ \\
  & Block-diagonal Clifford & $-$   & $\leq d^2$ & $ \leq d^3$ \\ \hline
  Arbitrary (integer ring) & Controlled-$Z_d$ &  $H_d$ & 4 & $ \sim 4 d^4$    \\
  & Controlled-$X_d$ & $S_d H_d^{-1} S_d$   &  $d$ & $\sim d^5 $ \\
  & Block-diagonal Clifford & $-$   & $\leq d^2$ & $\sim d^6 $
\end{tabular}
\caption{Comparison of qudit stabilizer state resources. We list the intrinsic gates associated with each resource, characterized by the respective entangling interaction, and the resulting total cost of implementing single-qudit unitaries measurement-based. For prime-power dimensions, we use the decomposition given in Eq. \eqref{eq:unitary-decomposition-primepower} without adapting the intrinsic gate. The qudits for the cluster state and light-shift gate resources are initialized in the equal superposition state $\ket{0_X}$, for controlled-$X$ gate resources in $S \ket{0_X}$, and for block-diagonal Clifford entangling gates in $D_{\Vec{\varphi}} \ket{0_X}$. Depending on the Pauli order $o_{G_I}^P$ and the dimension $d$, the number of measurements to implement an arbitrary single-qudit unitary is upper bounded by $o_{G_I}^P \cdot d$ for finite-field qudits of prime-power dimension and of the order $o_{G_I}^P \cdot d^4$ for integer-ring qudits of arbitrary dimension.
}
  \label{tab:overhead-singlequditunitary}
\end{table*}

In the following, we discuss previous resource state examples which can be mimicked by a qudit graph state with adjusted measurement bases. As discussed previously, the main idea is to effectively modify the intrinsic gate from $G_I$ to $D_2 G_I D_1$ with a lower Pauli order to reduce the overall measurement depth for the implementation of arbitrary single-qudit unitaries. In particular, we discuss that imitating both qutrit light-shift gate and controlled-$X$ gate resources allows for more efficient quantum information processing with the qutrit cluster state.

The intrinsic gate associated with the resource determines the effectiveness of the decomposition of arbitrary single-qudit unitaries into measurement patterns along one-dimensional resource chains, Fig. \ref{fig:setting} $(b)$. In particular, for prime-power-dimensional finite-field qudits, we group the $d^2$ generators of the unitary decomposition in Eq. \eqref{eq:unitary-decomposition} according to Eq. \eqref{eq:summarizing-generators}, allowing us to implement $d-1$ generators simultaneously. As a result, the implementation of any single-qudit unitary requires at most $d \cdot o_{G_I}^P$ measurements, which realize the pattern in Eq. \eqref{eq:unitary-decomposition-primepower}. For integer-ring qudits, each of the $d^2$ generators in Eq. \eqref{eq:unitary-decomposition} is implemented individually, and the required Clifford operations for that may need sub-quadratic many Hadamard and phase gates in the dimension $d$ \cite{Farinholt_2014}, so that in total the number of measurements is approximately at the order of $d^4 \cdot o_{G_I}^P$ (in fact, sub-quartic in $d$).

Hence, qubit light-shift gate resources, where $o_{U_2}^P=2$, require at most four measurements to implement an arbitrary single-qubit unitary, as for the cluster state resource. In contrast, for light-shift gate qutrits, the upper bound becomes $ 3 \cdot 3 = 9 $ due to $o_{U_3}^P= 3$, allowing for a more efficient measurement-based gate implementation via the decomposition in Eq. \eqref{eq:unitary-decomposition-primepower} than the qutrit cluster state resource which has $o_{H_3}=4$, so that up to twelve measurements may be necessary when not modifying the intrinsic gate. However, via adjusting measurement angles in two subsequent measurement rounds, the qutrit cluster state can mimic the more favorable intrinsic-gate structure $S_3^2 H_3 S_3^2$, reducing the Pauli order of the modified intrinsic gate and thereby improving the efficiency also for the qutrit cluster state.

For controlled-$X$ gate qubit and qutrit resources, the integer-ring and finite-field perspectives coincide: it then holds that
\begin{equation}
    (S H S)^2 = S H S^2 H S = S H Z H S = S X S = iXZ,
\end{equation}
so the Pauli order of the intrinsic gate is two. For qutrits, we have that $(S_3 H_3^{-1} S_3)^3 = I_3$, so the Pauli order is three. A qutrit graph state with edge weight $w = -1$ can mimic this controlled-$X$ gate resource since it has an intrinsic gate of $H_3 M(-1) = H_3^3 = H_3^{-1}$.

For finite-field ququarts, we have that $(S_4^F)^2 = Z_4(1) $ and $(H_4^F)^2 = I_4$ or $(H_4^F)^{-1} = H_4^F$ (as for any even prime-power dimension, see Appendix \ref{app:conjugation-cliffords}), so
\begin{equation}
    (S_4^F H_4^F S_4^F)^2 = S_4^F H_4^F Z_4(1) H_4^F S_4^F = -X_4(1) Z_4(1). 
\end{equation}
In contrast, for integer-ring ququarts, it is $(S_4 H_4^{-1} S_4)^4 = -X_4^2$, so the Pauli order is twice as large compared to finite-field ququarts. More generally, the Pauli order of controlled-$X$ gate resources can be shown to be $p$ for finite-field qudits and $d$ for integer-ring qudits (this follows from considering the associated symplectic matrices, which are introduced in Appendix \ref{app:universality-hadamard-diagonal}).
Therefore, it is favorable to work with the finite-field description whenever possible.

We summarize the unmodified intrinsic gates and their respective Pauli orders of cluster state, light-shift gate, controlled-$X$ gate, and block-diagonal Clifford gate resources in Tab. \ref{tab:overhead-singlequditunitary}. For odd prime-power dimensions, an upper bound on the Pauli order $o_{G_I}^P$ is known, which is linear in the dimension $d=p^m$, $p \neq 2$, namely $\max\{d+1,4p \}$ \cite{appleby2009propertiesextendedcliffordgroup}.
If no smaller upper bound for the Pauli order $o_{G_I}^P$ is available, we use $d^2$ since there are (up to phases) only $d^2$ distinct Pauli operators that a Clifford intrinsic gate $G_I$ can permute.

Note that in even prime-power dimensions, we have that $o_H = 2$ so that the upper bound on the number of measurements is always minimal (furthermore, in every dimension, it holds that $o_H \leq 4$, see Appendix \ref{app:integerring-cliffords}). Therefore, in even prime-power dimensions the cluster state is always optimal due to minimal Pauli order and we can only optimize the required measurement depths using the finite-field cluster state resource in dimensions with $p \neq 2$, as achieved with imitating the qutrit light-shift gate or controlled-$X$ gate resources for $p=3$.

\subsection{Non graph state-like resource states}
\label{subsec:non-graph-state-like}

In prime-power dimensions $p^m$ with $m>1$ the full unitary normalizer of the finite-field Pauli group is larger than the finite-field Clifford group, Sec. \ref{subsec:finite-field-Clifford}, and corresponds to the Clifford group on $m$ many $p$-dimensional qudits.

Resource states, characterized by an entangling normalizing gate outside of the finite-field Clifford group, can then not be understood via finite-field graph states and adjusted measurement bases. A counter-example for this was the ququart light-shift gate resource, Sec. \ref{sec:lightshift-resources}, where the entangling gate is a normalizing gate beyond the two-ququart finite-field Clifford group. Describing the ququart via two qubits, the associated intrinsic gate then corresponds to a two-qubit Clifford, Eq. \eqref{eq:lightshift-ququart-intrinsic}, instead of a single-ququart Clifford.

As the finite-field ququart cluster state, the ququart light-shift gate resource is optimal in the sense of achieving minimal Pauli order $o^P_{U_4} = 2$, so that at most eight measurements are required to decompose arbitrary single-ququart unitaries according to the measurement pattern in Eq. \eqref{eq:unitary-decomposition-primepower}.

In Appendix \ref{app:self-inverse-intrinsic}, we first show that the qudit cluster state or, more generally, an equal-weight qudit graph state with adjusted measurement bases cannot achieve a modified self-inverse intrinsic gate unless $p=2$. A natural question to then further ask is whether in dimensions with $p \neq 2$ alternative resource states admit self-inverse intrinsic gates, which also permit universal measurement-based quantum computing. We answer this question positively for $m$ even and negatively for $m$ odd in Appendix \ref{app:self-inverse-intrinsic}.

Furthermore, we explain in Appendix \ref{app:self-inverse-intrinsic} that is possible to realize such resources via diagonal entangling gates applied to qudits, initialized in $\ket{+}$, as in Sec. \ref{sec:diagonal-entangling-gates}. This is achieved by writing the angles $\theta_{ij}$ of the self-inverse intrinsic normalizing gate, which takes the form of Eq. \eqref{eq:native-operation}, into a diagonal entangling gate, Eq. \eqref{eq:diagonal-entangling}.

As an example, we construct in Appendix \ref{app:self-inverse-intrinsic} for $m=2$, the intrinsic gate
\begin{equation}
    (H \otimes H^{-1}) ((H \otimes H)CZ)^3.
    \label{eq:normalizing-self-inverse}
\end{equation}
This gate is self-inverse and leaves the Pauli group invariant while all $Z(z)$ with $0 \neq z \in \mathbb{F}_{p^2}$ are transformed to a non-zero $X$ argument, ensuring that universal measurement-based quantum computing is possible. A similar construction applies to all prime-power dimensions with an even exponent $m$ using the intrinsic gate
\begin{equation}
    \left((H \otimes H^{-1}) ((H \otimes H)CZ)^3 \right)^{\otimes \frac{m}{2}}.
\end{equation}
The smallest relevant example for $p$ odd is a nine-dimensional qudit, for which the matrices in Eq. \eqref{eq:normalizing-self-inverse} are qutrit matrices.

\section{Conclusion and outlook}
\label{sec:conclusion}

We have discussed generalized qudit stabilizer state resources beyond cluster states, allowing for exact universal measurement-based quantum computing. These resources, characterized by diagonal or block-diagonal Clifford entangling gates, offer greater flexibility in the entangling interaction, which may facilitate their preparation or availability in practice.
Furthermore, these alternative resources can potentially reduce the complexity of measurement-based quantum computing by rendering the decomposition of arbitrary single-qudit unitaries into measurement patterns more efficient.

We found that generalizing standard qubit measurement patterns to qudit cluster states is suboptimal in all but even prime-power dimensions. Moreover, we characterized which resource states can be mimicked by equal-weight qudit graph states via adjusting measurement bases for two subsequent measurements and discussed how, using these insights, the required overhead for measurement-based quantum computing with cluster states can be optimized. In certain dimensions, we found qudit resource states, which are optimal in terms of the required measurement depth and which can not be imitated by graph states. Overall, our results highlight that cluster states are not the unique optimal resources for measurement-based quantum computing, and that alternative qudit stabilizer states may offer both practical and theoretical advantages for scalable quantum computing.

The measurement-based implementation of arbitrary single-qudit unitaries was described in terms of an intrinsic single-qudit Clifford gate associated with each resource state, which is determined by the entangling gate and the initialization of the resource qudits. Herein, we have established a criterion for the intrinsic gate to obtain an exactly universal single-qudit gate set. Moreover, for prime-power dimensional qudits, we have shown that the decomposition of arbitrary single-qudit unitaries is efficiently possible, with an overhead that depends linearly on the dimension of the quantum system and the Pauli order of the intrinsic gate. Hence, the qudit cluster state can mimic more favorable intrinsic-gate structures to reduce measurement depth.

We found that all Clifford intrinsic gates satisfying the universality condition may be imitated by equal-weight qudit graph states and adjusted measurement bases. However, for prime-power-dimensional qudits, the full unitary normalizer of the Pauli group includes finite-field Clifford gates but goes beyond. This resulted in resource states that cannot be mimicked by qudit graph states for prime-power dimensions with an even exponent. An example for such a resource is the ququart light-shift gate state.

Since universal quantum computing also requires the ability to perform an entangling gate, we have discussed how the topology of a two-dimensional resource state can be designed to support arbitrary quantum information processing. For diagonal Clifford entangling gates, one can use the same two-dimensional lattices as for standard measurement-based quantum computing with cluster state resources. Instead, for block-diagonal Clifford entangling gates, we have proposed a different two-dimensional structure that relies on mediator qudits, allowing entangling or disentangling computational qudits. In addition, we have demonstrated that, as for cluster state resources, Clifford circuits can be implemented in a single time step on our generalized qudit resource states.

As an outlook, it would be relevant to consider stabilizer state resources characterized by arbitrary Clifford entangling gates and to explore whether non-Clifford unitary intrinsic gates (beyond normalizer gates) can yield universal resource states. In particular, this would require addressing the implementation of arbitrary single-qudit unitaries, dealing with measurement-induced random by-products to render the computation deterministic, and designing suitable two-dimensional resource state geometries for entangling gates. Alternative lattice geometries beyond the two-dimensional cluster state are also of interest. For instance, for qubits, graph states with hexagonal and triangular lattice geometries have been investigated \cite{UniversalResourcesMBQC, UniversalResourcesMBQC2}. Furthermore, graph state manipulation tools for non-commuting entangling gates, such as the block-diagonal gates studied here, present an interesting direction for future research.

\begin{acknowledgments}
We acknowledge support from the Austrian Research Promotion Agency (FFG) under Contract Number FO999914030‌ (Next Generation EU). In addition, this research was funded in whole or in part by the Austrian Science Fund (FWF) Grants 10.55776/P36009, 10.55776/P36010, and 10.55776/COE1. Finanziert von der Europäischen Union.‌
\end{acknowledgments}

\interlinepenalty = 10000

\bibliographystyle{apsrev4-2}
\bibliography{quditMBQC}

\begin{appendix}

\interlinepenalty = 0
\clearpage

\section{Pauli conjugation relations}
\label{app:conjugation-cliffords}

We describe explicit Pauli conjugation relations for integer-ring, Sec. \ref{app:integerring-cliffords}, and finite-field, \ref{app:conjugation-finitefield-cliffords}, Clifford gates.

\subsection{Integer-ring Clifford gates}
\label{app:integerring-cliffords}

The integer-ring Hadamard gate $H_d$ satisfies
\begin{equation}
    H_d X_d H_d^{\dagger} = Z_d, \quad H_d Z_d H_d^\dagger = X_d^{-1},
\end{equation}
where $d \in \mathbb{Z}_d$ is an arbitrary finite dimension.
The phase gate $S_d$ commutes with $Z_d$ and conjugates $S_d X_d S_d^\dagger = \tau_d X_d Z_d $ since for all $j \in \mathbb{Z}_d$
\begin{equation}
    S_d X_d \ket{j} = S_d \ket{j+1} = \tau_d^{j^2 + 2j + 1} \ket{j+1} = \tau_d X_d Z_d S_d \ket{j}.
\end{equation}
Therefore, the operator $Y_d = \tau_d X_d^{-1} Z_d^{-1}$ has the eigenstates $\ket{k_Y} = S_d \ket{k_X} = S_d H_d \ket{k_Z}$,
\begin{equation}
    \begin{aligned}
        & Y_d S_d H_d \ket{k_Z} = \tau_d X_d^{-1} S_d Z_d^{-1} H_d \ket{k_Z}
        \\ & = S_d X_d^{-1} H_d \ket{k_Z} = S_d H_d Z_d \ket{k_Z},
    \end{aligned}
\end{equation}
using that $S_d^\dagger X_d S_d = \tau_d Z_d^{-1} X_d$ (so $S_d^\dagger X_d^{-1} S_d = \tau_d^{-1} X_d^{-1} Z_d$) and $H_d^\dagger X_d^{-1} H_d = Z_d$. 

Furthermore, the square of the Hadamard gate equals the multiplication gate $M(-1)$ due to
\begin{equation}
\begin{aligned}
    & (H_d)^2 = \sum_{j,k,m,l} \omega^{jk+lm} \ket{j} \bra{k} \ket{l} \bra{m} = \sum_{j,m,k} \omega^{k(j+m)} \ket{j} \bra{m}
    \\ & = \sum_{j,m} \delta_{j+m,0} \ket{j} \bra{m} = \sum_j \ket{-j} \bra{j} = M(-1).
\end{aligned}
\label{eq:square-H}
\end{equation}
Squaring another time then yields the identity, so that $(H_d)^4 = I_d$ in all finite dimensions $d$.

The controlled-$Z_d$ gate conjugates
\begin{equation}
\begin{aligned}
    & CZ_d (X_d \otimes I) CZ_d^\dagger = \sum_{k,j} \ket{k+1} \bra{k} \otimes Z_d^{k+1} \ket{j} \bra{j} Z_d^{-k} 
    \\ & = \sum_{k,j} \ket{k+1} \bra{k} \otimes \omega^j \ket{j} \bra{j} = X_d \otimes Z_d.
\end{aligned}
\end{equation}

\subsection{Finite-field Clifford gates}
\label{app:conjugation-finitefield-cliffords}

Now we consider qudits of prime-power dimension, so $d=p^m$ with $p$ prime and $m$ being a positive integer.
The multiplication gate $M(\lambda)$ transforms
\begin{equation}
    Z(z) \mapsto Z(\lambda^{-1}z), \quad X(x) \mapsto X(\lambda x)
\end{equation}
with $x,z \in \mathbb{F}_{p^m}$ upon conjugation.
Therefore, it relates different $Z(z)$ gates and $X(x)$ gates to each other.

Conjugating the Pauli $Z(z)$ via the Hadamard gate results in
\begin{equation}
    \begin{aligned}
        & H_d^F Z(z) (H_d^F)^\dagger = H_d^F \sum_{w \in \mathbb{F}_d} \chi(zw) \ket{ w} \bra{w} (H_d^F)^\dagger
        \\ & = \sum_{u,w,v \in \mathbb{F}_{d}} \chi(zw+u w) \ket{u} \bra{v} \chi( - w v)
        \\ & =  \sum_{u,w,v \in \mathbb{F}_{d}} \chi(w(z+u - v)) \ket{u} \bra{v}
        \\& = \sum_{u,v \in \mathbb{F}_{d}} \delta_{z+u-v,0} \ket{u} \bra{v}
        = \sum_{u \in \mathbb{F}_{d}} \ket{v-z} \bra{v} = X(-z).
    \end{aligned}
\end{equation}
Here, we used that for $z+u-v = 0$, we have $\chi(0) = 1$, whereas for $z+u-v \neq 0$, $\sum_{w \in \mathbb{F}_{d}} \chi(w(z+u - v)) =0$. The latter can be seen from
\begin{equation}
    \begin{aligned}
        & \sum_{w \in \mathbb{F}_{d}} \chi(w(z+u - v)) = \sum_{w \in \mathbb{F}_{d}} \chi(w(z+u - v)+b)
        \\ & = \chi(b) \sum_{w \in \mathbb{F}_{d}} \chi(w(z+u - v)),
    \end{aligned}
\end{equation}
where $\chi(b) \neq 1$ for some appropriate $b \in \mathbb{F}_d$ and the shift map $X(b): \mathbb{F}_d \mapsto \mathbb{F}_d, g \mapsto g + b$ with $b \in \mathbb{F}_d$ is a bijection.

In turn, conjugating $X(x)$ via the finite-field Hadamard gate $H_d^F$, results in
\begin{equation}
    \begin{aligned}
        & H_d^F X(x) (H_d^F)^\dagger = H_d^F \sum_{w \in \mathbb{F}_d} \ket{ w+x} \bra{w} (H_d^F)^\dagger
        \\ & = \sum_{u,w,v \in \mathbb{F}_{d}} \chi(u (w+x)) \ket{u} \bra{v} \chi( - w v)
        \\ & =  \sum_{u,w,v \in \mathbb{F}_{d}} \chi(w(u-v)) \chi(ux) \ket{u} \bra{v}
        \\& = \sum_{u,v \in \mathbb{F}_{d}} \delta_{u,v} \chi(ux) \ket{u} \bra{v}
        = \sum_{u \in \mathbb{F}_{d}} \chi(ux) \ket{u} \bra{u} = Z(x).
    \end{aligned}
\end{equation}

Similarly to before, we see that $(H_d^F)^2 = M(-1)$ and $(H_d^F)^4 = I_d$. For $p=2$, it even holds that $(H_d^F)^2 = I_d$ since $M(-1) = M(1)=I_d$.

The phase gate commutes with $Z(z)$, whereas for $S(\lambda) X(x) S(\lambda)^{-1}$ with $p \neq 2$, we obtain
\begin{equation}
\begin{aligned}
    & \sum_{u \in \mathbb{F}_{p^m}} \chi(2^{-1} \lambda (u+x)^2 - 2^{-1} \lambda u^2) \ket{u+x} \bra{u}
    \\ & = \sum_{u \in \mathbb{F}_{p^m}} \chi(2^{-1} \lambda (2ux + x^2 )) \ket{u+x} \bra{u}
    \\ & =  \chi(2^{-1} \lambda x^2 ) \sum_{u \in \mathbb{F}_{p^m}} \chi( \lambda ux) \ket{u+x} \bra{u}
    \\ & = \chi(2^{-1} \lambda x^2 ) X(x) Z(\lambda x),
\end{aligned}
\end{equation}
and for $p=2$
\begin{equation}
\begin{aligned}
    & S X(x) S^{-1} = \sum_{u \in \mathbb{F}_{p^m}} \chi_4((u+x)^2-u^2 ) \ket{u+x} \bra{u}
    \\ &= \sum_{u \in \mathbb{F}_{p^m}} \chi_4(x^2 + 2ux) \ket{u+x} \bra{u}
     \\&  = \chi_4(x^2) \sum_{u \in \mathbb{F}_{p^m}} \chi(ux) \ket{u+x} \bra{u} = \chi_4(x^2) X(x) Z(x).
     \label{eq:evenS-conjugation}
\end{aligned}
\end{equation}
Furthermore, for even prime-powers, $S^2 = Z(1)$ since $\chi_4(2x^2) = \chi(x^2) = \chi(x)$.

Note that for odd prime dimensions $d=p$, the finite-field phase gate coincides with the integer-ring phase gate due to
\begin{equation}
    \begin{aligned}
        & S_p^F (1) = \sum_{x \in \mathbb{Z}_p} \omega_p^{ 2^{-1} x^2 } \ket{x} \bra{x} = \sum_{x \in \mathbb{Z}_p} e^{ \frac{2 \pi i}{p} \frac{p+1}{2} x^2 } \ket{x} \bra{x}
        \\ & = \sum_{x \in \mathbb{Z}_p} (-e^{ \frac{ \pi i}{p}  })^{x^2} \ket{x} \bra{x} = \sum_{x \in \mathbb{Z}_p} \tau_p^{x^2 } \ket{x} \bra{x} = S_p,
    \end{aligned}
\end{equation}
where we have used the fact that the inverse of two can always be expressed as $2^{-1} = \frac{p+1}{2}$ in the finite field $\mathbb{Z}_p$.

The controlled-$Z$ gate $CZ$ conjugates $X(x) \otimes I_d$ to
\begin{equation}
\begin{aligned}
    \sum_{k,j} \ket{k+x} \bra{k} \otimes Z(k+x) \ket{j} \bra{j} Z(-k) 
    = X(x) \otimes Z(x).
\end{aligned}
\end{equation}

\section{Finite-field ququarts}
\label{app:finite-field-ququarts}

Ququarts are the smallest non-prime-dimensional system, where we can work with the prime-power formalism instead of the integer ring $\mathbb{Z}_4$. For this, we consider the field
\begin{equation}
    \mathbb{F}_4 = \mathbb{F}_{2^2} \cong \mathbb{F}_2[\xi]/ \langle f(\xi) \rangle,
\end{equation}
where $\mathbb{F}_2[\xi]$ is a polynomial ring in the variable $\xi$ with coefficients from $\mathbb{F}_2 = \mathbb{Z}_2$ and $f(\xi)$ is an irreducible polynomial of degree two (meaning that $f(\xi)$ cannot be factored into non-constant polynomials). The basis elements of the field can then be identified with $\ket{0}$, $ \ket{1}$, $\ket{\xi}$, $\ket{1+\xi}$ using the irreducible polynomial $f(\xi) = \xi^2 + \xi + 1 \equiv 0$ (it has no zeros in $\mathbb{Z}_2 = \{0,1 \}$, so it cannot be factored).

The finite-field Pauli gates act according to \cite{heinrich2021stabiliser}
\begin{equation}
    X_4(x) = \sum_{j \in \mathbb{F}_4} \ket{j + x} \bra{j}, \quad Z_4(z) = \sum_{j \in \mathbb{F}_4} (-1)^{\tr(zj)} \ket{j} \bra{j},
\end{equation}
where $x,z \in \mathbb{F}_4$ and the finite field trace of $j \in \mathbb{F}_4$ is $\tr: \mathbb{F}_{4} \mapsto \mathbb{Z}_2$, $j \mapsto j + j^2$,
\begin{equation}
    \begin{aligned}
        & \tr(j) = \tr(a + b \xi) = (a + b \xi) + (a + b \xi)^2
        \\ &= a + b \xi + a^2 + b^2 \xi^2 = b (\xi^2 + \xi) = b,
    \end{aligned}
\end{equation}
using that in $\mathbb{F}_2$, we have $a^2 = a$ and $2a =0$ as well as $f(\xi) \equiv 0$.

The Pauli gates can be related to each other via the multiplication gate $M(\lambda)$ since
\begin{equation}
    \begin{aligned}
    & M(\lambda) X(x) M(\lambda^{-1}) = \sum_{j \in \mathbb{F}_4} \ket{\lambda x + \lambda j} \bra{\lambda j} = X(\lambda x)
    \\ & M(\lambda) Z(z) M(\lambda^{-1}) = \sum_{j \in \mathbb{F}_4} (-1)^{\tr(jz)} \ket{\lambda j} \bra{\lambda j} = Z(\lambda^{-1} z).
    \end{aligned}
\end{equation}
The two non-trivial ququart multiplication gates are the ones corresponding to multiplication with $\xi$ and $1+\xi$ (which are also the inverses of each other) and are given by
\begin{equation}
    M(\xi) = \left( \begin{array}{cccc}
         1 & 0& 0 & 0 \\
         0 & 0 & 0 & 1 \\
         0 & 1 & 0 & 0 \\
         0 & 0 & 1 & 0 \\
    \end{array} \right),
    \quad
    M(\xi+1) = \left( \begin{array}{cccc}
         1 & 0& 0& 0 \\
         0 & 0 & 1 & 0 \\
         0 & 0 & 0 & 1 \\
         0 & 1 & 0 & 0 \\
    \end{array} \right).
\end{equation}
One can also check that $M(\xi) M(1+\xi) = I_4$ as well as $M(1+\xi) = M(\xi^2) = M(\xi) M(\xi)$.

The explicit Pauli matrices are given by
\begin{equation}
    \begin{aligned}
        & Z_4(1) = \left( \begin{array}{cccc}
         1& 0 & 0 &0  \\
         0& 1 & 0 &0 \\
         0& 0 & -1 &0 \\
         0& 0 & 0 & -1
    \end{array} \right), \quad
    Z_4(\xi) = \left( \begin{array}{cccc}
         1& 0 & 0 &0  \\
         0& -1 & 0 &0 \\
         0& 0 & -1 & 0 \\
         0& 0 & 0 & 1
    \end{array} \right)
    \end{aligned}
\end{equation}
and
\begin{equation}
    \begin{aligned}
        & X_4(1) = \left( \begin{array}{cccc}
         0& 1 & 0 &0  \\
         1& 0 & 0 &0 \\
         0& 0 & 0 & 1 \\
         0& 0 & 1 & 0
    \end{array} \right), \quad
    X_4(\xi) = \left( \begin{array}{cccc}
         0& 0 & 1 & 0  \\
         0 & 0 & 0 & 1 \\
         1 & 0 & 0 & 0 \\
         0 & 1 & 0 & 0
    \end{array} \right).
    \end{aligned}
\end{equation}
The remaining Pauli matrices are then specified by $Z_4(1 + \xi) = Z_4(1) Z_4(\xi)$ and $X_4(1 + \xi) = X_4(1) X_4(\xi)$.

The Hadamard gate \cite{heinrich2021stabiliser} can be expressed via
\begin{equation}
    \begin{aligned}
         H_4^F = \frac{1}{2} \sum_{k,j} (-1)^{\tr(jk)} \ket{j} \bra{k} = \frac{1}{2} \left( \begin{array}{cccc}
         1& 1 & 1 & 1  \\
         1& 1 & -1 & -1 \\
         1& -1 & -1 & 1 \\
         1& -1 & 1 & -1
    \end{array} \right).
    \end{aligned}
\end{equation}

To define the $S_4^F$ gate, we have to consider the Galois extension ring $\mathbb{GR}(4,m) = \mathbb{Z}_4[\xi] / \langle f(\xi) \rangle$ of the $\mathbb{Z}_4$ ring with $4^m$ elements, where $m$ is the degree of the polynomial $f(\xi)$, irreducible in $\mathbb{Z}_2$, in our case $m=2$. The ring elements are of the type $a_0 + a_1 \xi + \hdots + a_{m-1} \xi^{m-1}$ with $a_0,\hdots, a_{m-1} \in \mathbb{Z}_4$ and we can take the irreducible polynomial
\begin{equation}
    f(\xi) = \xi^2 + 3 \xi + 3 = \xi^2 - \xi - 1 \equiv 0
\end{equation}
since it has no roots in $\mathbb{Z}_2$, so that $ \xi^2 \equiv \xi + 1$. The finite field $\mathbb{F}_{2^m}$ is embedded in the above extension ring and an isomorphism obtained via $\mathbb{GR}(4,m) / \mathbb{Z}_2 \cong \mathbb{F}_{2^m}$. We can view $\mathbb{GR}(4,m)$ also as an $\mathbb{Z}_4$-module $R_{4^m}$ with elements $(a_0,\hdots , a_{m-1})$ and scalar multiplication being performed with $\mathbb{Z}_4$. A module is a generalization of the vector space concept, where scalar multiplication is performed with elements of a ring instead of a field.
Now the trace is taken of the linear map on $R_{4^m}$, denoted via $\tr_4$, which acts as $j \mapsto tj$,
\begin{equation}
    \tr_4: \mathbb{GR}(4,m) \mapsto \mathbb{Z}_4.
\end{equation}
We then have for $\tr_4(j^2)$ and $j \in \mathbb{F}_{2^m} \subset \mathbb{GR}(4,m)$
\begin{equation}
    j \in \mathbb{GR}(4,m) \mapsto j^2 \in \mathbb{GR}(4,m) \mapsto \tr_4(j^2) \in \mathbb{Z}_4.
\end{equation}

More concretely, representing polynomials $a_0 + a_1 \xi$ as elements $(a_0,a_1)$ of the module $R_{4^2}$ and permitting coefficients in $\mathbb{Z}_4$, multiplication with $0^2 = 0 \in \mathbb{F}_{4}$ is described by the zero matrix with trace 0, multiplication with $1^2 = 1$ via the identity matrix $I_2$ with trace 2, multiplication with $\xi^2 = 1 + \xi$ via
$\left( \begin{array}{cc}
     1 & 1 \\
     1 & 2
\end{array} \right)$ (since $(1+\xi)\xi = \xi + \xi^2 = 1 + 2\xi$) with trace 3
 and multiplication with $(1+\xi)^2 = 1 + 2\xi + \xi^2 = 2 + 3\xi$ via $\left( \begin{array}{cc}
     2 & 3 \\
     3 & 1
\end{array} \right)$ (since $(2 + 3\xi) \xi = 2 \xi + 3(1+\xi) = 3 + \xi$) with trace 3.

The phase gate matrix then becomes:
\begin{equation}
    S_4^F = \sum_{j \in \mathbb{F}_{4}} i^{\tr_4(j^2)} \ket{j} \bra{j} = \left( \begin{array}{cccc}
         1& 0 & 0 &0  \\
         0& -1 & 0 &0 \\
         0& 0 & -i &0 \\
         0& 0 & 0 & -i
    \end{array} \right).
\end{equation}
We see that $(S_4^F)^4 = I_4$ and $(S_4^F)^2 = Z_4(1)$.
The $S_4^F$ gate trivially commutes with $Z(z)$ gates, whereas conjugation of $X_4(x)$ yields according to Eq. \eqref{eq:evenS-conjugation} $ \chi_4(x^2) X(x) Z(x)$.

\section{Universality of resource stabilizer states}

For universal measurement-based quantum computing, we need to ensure that the measurement-based gate set can realize any unitary operation. This implies showing that the intrinsic gate, supplemented by diagonal phase gates, allows the implementation of any single-qudit unitary. We demonstrate this in Appendix \ref{app:universality-finite-field} for finite-field qudits and in Appendix \ref{app:universality-integer-ring} for integer-ring qudits. In both cases, single-qudit unitaries are realized on one-dimensional resource state chains such as in Fig. \ref{fig:setting} $(b)$. In addition, we discuss the decomposition of any single-qudit unitary into measurement patterns and, in particular, the number of required measurements.

Furthermore, once we can implement single-qudit unitaries, any two-qudit entangling gate is sufficient to obtain a universal gate set \cite{QuditComputing}. The required two-dimensional resource state geometry to allow both for single-qudit unitary gate implementation and a two-qudit entangling gate implementation in a measurement-based fashion is discussed for diagonal Clifford entangling gates in Appendix \ref{app:LC-equivalent-cluster-state} and for block-diagonal Clifford entangling gates in Appendix \ref{app:2Dresource-blockdiagonal}. Finally, we discuss in Appendix \ref{app:clifford-circuits} that Clifford circuits can be realized in a single time step for our generalized qudit stabilizer state resources.

\subsection{Implementation of arbitrary single-qudit unitaries for finite-field qudits}
\label{app:universality-finite-field}

To show that any single-qudit unitary can be implemented on finite-field qudits, we generalize the proof in Ref. \cite{Clark_2006}. There, the author shows that the Hadamard gate, supplemented with diagonal phase gates, generates arbitrary single-qudit unitaries for prime-dimensional qudits. We show that replacing the Hadamard gate with a Clifford intrinsic gate, associated with our resource stabilizer state, that conjugates
\begin{equation}
    Z(z) \mapsto Z(a (z)) X(b(z)), \quad b(z) \neq 0 \quad \forall z \neq 0
\end{equation}
allows for the same argument and that this argument also holds in prime-power dimensions.

\subsubsection{Computational universality of intrinsic gate and diagonal phase gates}

Any single-qudit unitary $U$ can be rewritten with a suitable Hermitian matrix $H$ via $U=e^{iH}$. For instance, we can decompose $U = V D V^{\dagger}$, where
\begin{equation}
    D = \textnormal{diag} (e^{i\theta_1},\dots,e^{i\theta_d})
\end{equation}
is a diagonal matrix with the eigenvalues $\{e^{i \theta_j}\}_{j=1}^{d}$ of $U$, $\theta_j \in \mathbb{R}$, on its diagonal, and set
\begin{equation}
    H \coloneq \frac{1}{i} V \ln(D) V^\dagger = V \textnormal{diag} (\theta_1,\dots,\theta_d) V^\dagger.
\end{equation}
Given a Hermitian basis $N_i$ with $i \in \mathbb{Z}_{d^2},$ (for qubits, so $d=2$, one could, for instance, choose $\{I,X,Z, iZX \}$) we can expand $H$ in this basis,
\begin{equation}
    U = e^{iH} = e^{i\sum_{k=1}^{d^2} \alpha_k N_k} = \prod_{k=1}^{d^2} e^{i \beta_k N_k}
\end{equation}
for some real parameters $\alpha_k,\beta_k$ (the order of the product can be chosen as desired, however, this generally affects $\beta_k$) \cite{Clark_2006, Zhou_2003, LieAlgebraExponentialProducts}.

If the dimension $d$ is an odd prime, we can construct a Hermitian basis via projectors onto Pauli eigenstates \cite{Clark_2006}. This also works for prime-power dimensions and relies on the fact that there exist $d+1$ so-called mutually unbiased bases via Pauli eigenstates \cite{MUBRef21, DURT_2010}.
A set of bases is said to be mutually unbiased if each basis is orthonormal and the modulus square of the inner product of two vectors $\ket{v}, \ket{w}$ from two different bases is $1/d = | \bra{v}\ket{w} |^2$ \cite{MUBRef21}.

Hence, for each of the $d+1$ Paulis $P$, we consider its eigenbasis $\{ \ket{k_P} \}_{k \in \mathbb{Z}_{d}}$ and form the Hermitian projectors $ \{ \ket{k_P} \bra{k_P} \}_{k \in \mathbb{Z}_{d-1}}$, omitting the first eigenvector in all bases. For an appropriate choice of the Paulis $P$, these eigenbases are mutually unbiased and one can then show that the resulting $(d+1)(d-1)=d^2 -1$ Hermitian operators are linearly independent \cite{Clark_2006}. If we take for one of the Paulis $P$ all eigenvectors, we include the identity $I_d$ and have $d^2$ Hermitian linearly independent operators. However, since the identity only changes the global phase of the unitary $U$, constructing $I_d$ is irrelevant.

More explicitly, the $d+1$ Paulis whose eigenbases we pick are $Z(1)$ and \cite{MUBRef21}
\begin{equation}
    \{ X(1) Z(a) \mid a \in \mathbb{F}_{d} \}.
\end{equation}
For prime dimensions $d$, this corresponds to the set of Pauli operators $\{X,Z, XZ, X Z^2, \hdots, XZ^{d-1} \}$.
For distinct $b_1 \neq b_2$ in $\mathbb{F}_{d}$, we have due to Eq. \eqref{eq:commutation-finitefield-Paulis} that
\begin{equation}
\begin{aligned}
    & X(b_1) Z(b_1 a) X(b_2) Z(b_2 a) = X(b_2) Z(b_2 a) X(b_1) Z(b_1 a),
\end{aligned}
\end{equation}
so that two operators in $ \{ X(b) Z(ba) \mid b \in \mathbb{F}_{d} \}$ commute and posses a joint eigenbases. Hence, the eigenbasis of $X(1)Z(a)$ represents the eigenbasis of all operators within the set $ \{ X(b) Z(ba) \mid b \in \mathbb{F}_{d} \}$.

The simplest parts of the Hermitian basis are the $d$ projectors onto the eigenvectors of $Z$, $\{ \ket{k_Z} \bra{k_Z} \}_{k \in \mathbb{Z}_d}$, whose exponentiation according to Eq. \eqref{eq:unitary-decomposition} corresponds to implementing a diagonal matrix.
If we have conjugation with the Hadamard gate $H_d^F$ and with the phase gate $S_d^F(\lambda)$ at our disposal, the Hadamard gate conjugates the $Z$ eigenbasis to the $X$ eigenbasis, whereas conjugation with the phase gate allows us to cycle through the remaining Pauli eigenbases of Eq. \eqref{eq:disjoint-eigensets} due to $X(1) \mapsto X(1)Z(\lambda)$. Here, we use that if $\ket{\psi}$ is an eigenvector of a Pauli $P$, then $ U \ket{\psi} \propto U P \ket{\psi} = U P U^\dagger U \ket{\psi} $, so $U \ket{\psi}$ is an eigenvector of the conjugated Pauli operator $U P U^\dagger$. Conjugation with $S_d^F(\lambda) H_d^F$ then allows us to obtain any eigenbasis of the Paulis in Eq. \eqref{eq:disjoint-eigensets}.

For even prime-power dimensions, the gate $S_d^F(\lambda)$ is replaced with $M(\lambda) S_d^F M(\lambda^{-1})$ with $0 \neq \lambda \in \mathbb{F}_{2^m}$, which maps $X(1)Z(a)$ to $X(1) Z(a + \lambda^{-2})$.
Note that in $\mathbb{F}_{2^m}$ the map $\lambda \mapsto \lambda^2$ is a bijection due to
\begin{equation}
    \begin{aligned}
        & (a_0 + a_1 \xi + \hdots + a_{m-1} \xi^{m-1})^2
        \\ & = a_0 + a_1 \xi^2 + \hdots + a_{m-1} (\xi^{m-1})^2.
    \end{aligned}
\end{equation}
Therefore, we can again cycle through all Pauli eigenbases. We denote both variants of the phase gate (for even and odd prime-power dimensions) with $S(\lambda)$.

The measurement-based gate set at our disposal is $\{ G_I, D_{\Vec{\phi}} \mid \phi \in \mathbb{R}^d \}$. Requiring that $G_I$ is a Clifford gate that does not map the computational basis onto itself, so it conjugates $Z(z) \mapsto Z(a(z)) X(b(z))$ with $a(z), b(z) \in \mathbb{F}_d$ and $b(z) \neq 0$ (up to a phase) for all $0 \neq z \in \mathbb{F}_d$, we argue in the following that $G_I$ can replace the role of $H_d^F$.

First of all, having $b \neq 0 $, the eigenbases of $Z(a) X(b)$ and $X(1) Z(a b^{-1})$ are equal.
The phase gate $S(\lambda)$ then maps $X(1) Z(a b^{-1}) \mapsto X(1) Z(a b^{-1} -\lambda)$ (or to $X(1) Z(a b^{-1} -\lambda^2)$ for $p=2$, respectively), allowing us to obtain all distinct eigenbases. Thus, once we have any of the non-$Z$ eigenbases, provided by conjugation with $G_I$, the diagonal gates $S(\lambda)$ allow us to obtain the remaining eigenbases.

To conjugate with $G_I$ measurement-based, we need to apply $G_I^\dagger = G_I^{-1}$. Since $G_I$ is Clifford, so a permutation of Pauli group elements, it automatically has finite order (there are only finitely many ways to permute a finite set), so that there exists an integer $o_{G_I}$ with $(G_I)^{o_{G_I}} = I_d$. Then, the inverse of $G_I$ can be realized via $(G_I)^{o_{G_I} - 1} = G_I^\dagger$. Note that only the Pauli order is relevant, so $o_{G_I}^P$ such that $(G_I)^{o_{G_I}^P}$ is a Pauli operator since we always compute up to Pauli by-products.

For non-prime-power dimensions, the existence of maximal sets of mutually unbiased bases is an open problem, even for the most studied case of $d=6$ \cite{DURT_2010}. It even seems unlikely that more than three mutually unbiased bases exist for $d=6$ \cite{DURT_2010}. Since $d+1$ is always the maximum number of mutually unbiased bases over $\mathbb{C}^d$, finding $d+1$ bases tells us that we have found a maximal set \cite{MUBRef21}.

In the next section, we explicitly decompose an arbitrary single-qudit unitary into a measurement pattern, deriving an upper bound on the number of required measurements for its implementation.

\subsubsection{Upper bound for measurement pattern that implements an arbitrary single-qudit unitary}

Any single-qudit measurement on our resource stabilizer state implements the gate $G_I D_{\Vec{\phi}}$ (up to a Pauli by-product), where $G_I$ is the intrinsic gate and $D_{\Vec{\phi}}$ is an arbitrary diagonal unitary.

As mentioned previously, we assume that the intrinsic gate $G_I$ is a Clifford operation that maps the $Z(1)$ basis to some $ Z(a) X(b)$ with $b \neq 0$. Then $\{ G_I \ket{k_Z}  \}_k$ directly provides us with a basis of Pauli eigenstates. Furthermore, since we construct the Hermitian basis from projectors on Pauli eigenstates, within one set, the eigenvectors $\{ \ket{k_P} \}_k$ are orthogonal. Hence, the projectors commute and one can group
\begin{equation}
    \prod_k e^{i \alpha_k \ket{k_P} \bra{k_P}} =  e^{i \sum_k \alpha_k \ket{k_P} \bra{k_P}}
\end{equation}
together. The simplest parts of the Hermitian basis are the $d$ projectors onto the eigenvectors of $Z$, $\{ \ket{k_Z} \bra{k_Z} \}_{k \in \mathbb{Z}_d}$, which can be grouped together for the unitary decomposition of Eq. \eqref{eq:unitary-decomposition} into a single diagonal matrix
\begin{equation}
    \prod_{k=1}^d e^{i \alpha_k \ket{k_Z} \bra{k_Z}} = e^{i \sum_{k=1}^d \alpha_k \ket{k_Z} \bra{k_Z}} = D_{\Vec{\alpha}}.
\end{equation}
For instance, implementing a qubit $Z$ rotations corresponds to the choice $\alpha_0 = \alpha$, $\alpha_1 = - \alpha$,
\begin{equation}
    e^{i \sum_k \alpha_k \ket{k_Z} \bra{k_Z}} = e^{i (\alpha \ket{0} \bra{0}- \alpha \ket{1} \bra{1})} =e^{i \alpha Z}.
\end{equation}

Conjugation with $S(\lambda) G_I$ transforms a previously diagonal matrix into
\begin{equation}
    \begin{aligned}
        & S(\lambda) G_I D_{\Vec{\alpha}} G_I^\dagger S(\lambda)^\dagger = \prod_{k=1}^d e^{i \alpha_k S(\lambda) G_I \ket{k_Z} \bra{k_Z} G_I^\dagger S(\lambda)^\dagger},
    \end{aligned}
\end{equation}
where $\{ S(\lambda) G_I \ket{k_Z} \bra{k_Z} G_I^\dagger S(\lambda)^\dagger \}_k$ corresponds to the non-$Z$ Pauli eigenstate projectors of Eq. \eqref{eq:disjoint-eigensets} if we cycle through all $\lambda \in \mathbb{F}_d$ (note that $S(0)$ is formally undefined and should be interpreted as $I_d$ in the context here).
As before, for odd prime-power dimensions, $S(\lambda)$ is the phase gate $S_d^F(\lambda)$ from Eq. \eqref{eq:finitefield-phasegate-odd} and, for even prime-power dimensions, it is $M(\lambda) S_d^F M(\lambda^{-1})$ with $S_d^F$ from Eq. \eqref{eq:finitefield-phasegate-even}. Note that conjugation cancels potential phases introduced by $S(\lambda) G_I$.

The diagonal Hermitian basis elements $\ket{k_Z} \bra{k_Z}$, or the generators $\prod_k e^{i \alpha_k \ket{k_Z} \bra{k_Z}}$, respectively, correspond to the implementation of a diagonal matrix $D_{\Vec{\alpha}}$ and require $o_{G_I}^P$ measurements (to remove $G_I$, additional $o_{G_I}-1$ measurements are necessary in addition to one measurement realizing $G_I D_{\Vec{\alpha}}$). Each set of $d-1$ non-diagonal generators $ \prod_k e^{i \alpha_k \ket{k_P} \bra{k_P}} $ is obtainable in terms of $2o_{G_I}^P$ measurements up to Paulis via
\begin{equation}
\begin{aligned}
    & \prod_k e^{i \alpha_k \ket{k_P} \bra{k_P}} = e^{i \sum_k \alpha_k S(\lambda) G_I \ket{k_Z} \bra{k_Z} G_I^\dagger S(\lambda)^\dagger}
    \\ & = S(\lambda) G_I e^{i \sum_k \alpha_k \ket{k_Z} \bra{k_Z} } G_I^\dagger  S(\lambda)^\dagger
    \\ & = G_I^{o_{G_I}-1} (G_I  S(\lambda)) (G_I D_{\Vec{\alpha}}) G_I^{(o_{G_I}-2)} (G_I  S(\lambda)^\dagger)
\end{aligned}
\end{equation}
for an appropriate $\lambda \in \mathbb{F}_d$.

Any single-qudit unitary can then be decomposed into
\begin{equation}
\begin{aligned}
    G_I D_{\Vec{\gamma}} G_I^\dagger \left( \prod_{\lambda \in \mathbb{F}_d \backslash \{ 0\}} S(\lambda) G_I D_{\Vec{\beta}(\lambda)} G_I^\dagger S(\lambda)^\dagger \right) D_{\Vec{\alpha}},
\end{aligned}
\end{equation}
where $D_{\Vec{\alpha}} = e^{i \sum_k \alpha_k \ket{k_Z} \bra{k_Z}} $ with real parameters $\{ \alpha_k \}_k$ and analogous for the $d-1$ diagonal matrices $D_{\Vec{\beta}(\lambda)}$ and for $D_{\Vec{\gamma}}$.

This decomposition immediately shows that we require at most $d o_{G_I}$ measurements to implement an arbitrary single-qudit unitary due to
\begin{equation}
    (d-1)(o_{G_I} -1 + 1) + o_{G_I} -1 + 1 = d o_{G_I},
\end{equation}
where each $G_I^\dagger D_{\Vec{\phi}}$ operation requires $o_{G_I} - 1$ many measurements, each $G_I D_{\Vec{\phi}}$ a single measurement, and all subsequent diagonal gates are summarized into a single diagonal gate.
For cluster state qubits, so $d=2$ and $G_I = H$, this reproduces the known result of requiring at most four measurements for the measurement-based implementation of a single-qubit unitary (although this result is typically derived from the Euler decomposition of single-qubit unitaries).
For instance, for cluster state qubits, we could decompose any single-qubit unitary with appropriate $\alpha, \beta, \gamma \in \mathbb{R}$ according to
\begin{equation}
\begin{aligned}
    & e^{i \gamma H \ket{0_Z} \bra{0_Z} H } e^{i \beta S H \ket{0_Z} \bra{0_Z} H S^\dagger }  e^{i \alpha \ket{0} \bra{0}}
    \\ & = (H D_{\gamma} ) (H S)(H D_{\beta}) ( H S^\dagger D_{\alpha}).
\end{aligned}
\end{equation}

\subsection{Implementation of arbitrary single-qudit unitaries for integer-ring qudits}
\label{app:universality-integer-ring}

For integer-ring qudits of arbitrary finite dimension $d$, we show in the following that the Hadamard gate and diagonal gates allow us to perform any single-qudit unitary. Next, we prove that the intrinsic gate and phase gates can produce the Hadamard gate, provided that the Clifford intrinsic gate conjugates the computational basis $Z_d \mapsto Z_d^a X_d^b$ with $b \in \mathbb{Z}_d^*$ being invertible. For simplicity, we omit the index $d$ and write $X$ and $Z$ for the integer-ring Paulis in the next section.

\subsubsection{Universality of Hadamard and diagonal gates to generate arbitrary single-qudit unitaries}
\label{app:universality-hadamard-diagonal}

A Clifford operation maps $Z \mapsto Z^a X^b$ and $X \mapsto Z^c X^d$ (up to phases) with $ad - bc = 1$ to preserve the commutation relation of Eq. \eqref{eq:commutation-modular-Paulis} for the conjugated Pauli operators.
Thus, we can represent any Clifford gate, such as the intrinsic gate $G_I$ and its inverse $G_I^{-1}$, via symplectic matrices (so matrices with determinant one)
\begin{equation}
    G_I \mapsto \left( \begin{array}{cc}
         a & c  \\
         b & d
    \end{array} \right),
    \quad
    G_I^{-1} \mapsto \left( \begin{array}{cc}
         d & - c  \\
         - b & a
    \end{array} \right),
    \label{eq:intrinsicgate-symplectic}
\end{equation}
where we associate the Pauli $Z^{k_1} X^{k_2}$ with the column vector $ (k_1,k_2)^T \in (\mathbb{Z}_d)^2$.

In Ref. \cite{Farinholt_2014}, the author shows that the phase and Hadamard gates are a necessary and sufficient set of gates to generate (up to global phases) the entire single-qudit Clifford group in any finite dimension by explicit decomposition of an arbitrary symplectic matrix.

Then, one can obtain a Hermitian basis from the unitary operator basis $\{ Z^a X^b \mid a,b \in \mathbb{Z}_d \}$ \cite{Zhou_2003, Asadian_2016}, which in turn can be expressed via Pauli eigenstate projectors.
For instance, in Ref. \cite{Asadian_2016}, the authors construct the Hermitian operators $N(a,b) = c e^{- \frac{i \pi  ab }{d}} Z^{a} X^{b} + c^* e^{\frac{i \pi  ab }{d}} Z^{-b} X^{-a}$ ensuring with the appropriate choice of $c = \frac{1 \pm i}{\sqrt{2}} \in \mathbb{C} $ linear independence (even orthogonality with respect to the Hilbert-Schmidt inner product) \cite{Asadian_2016}. 

For integer-ring qudits of non-prime dimension, the eigenvectors of the Pauli operators in the set $\{ Z, X, XZ, \hdots , X Z^{d-1} \}$ do not represent all distinct Pauli eigenstates anymore. For instance, if $X^{k_1} Z^{k_2}$ with $k_1$ and $k_2$ being non-invertible but unequal zero (which does not happen in the finite-field case), cannot be related to one of the Pauli operators in the above set via an appropriate power.
In general, we are only interested in eigenbases of Paulis $Z^{k_1} X^{k_2}$ with the greatest common divisor $\gcd(k_1,k_2) = 1$ since otherwise we can relate it to the power of another Pauli operator which has the same eigenbasis.
If $\gcd(k_1, k_2) = 1$, we know due to the extended Euclidean algorithm (or Bézout's identity) that there exist unique integers $u,v \in \mathbb{Z_d}$ such that $1 = u k_1 + v k_2$. Therefore, a Clifford operator $A$ exists which maps $Z$, represented by $(1,0)^T$, onto $Z^{k_1} X^{k_2}$, namely $A$ corresponds to the symplectic transformation
\begin{equation}
    A \mapsto \left( \begin{array}{cc}
     k_1 & -v \\
     k_2 & u
\end{array} \right).
\label{eq:symplectic-any-pauli-basis}
\end{equation}
Thus, the Hadamard gate $H_d$ and phase gate $S_d$ generating all symplectic matrices implies that we can obtain all distinct Pauli eigenbases projectors $\ket{k_P} \bra{k_P}$ via conjugation of the computational basis.
Hence, we show in the following how to obtain the Hadamard gate symplectic matrix,
$\left( \begin{array}{cc}
         0 & 1  \\
         -1 & 0
    \end{array} \right)$, from the Clifford gate $G_I$ and $S^l$ gates, the latter being associated with the symplectic matrices $ \left( \begin{array}{cc}
     1& l \\
     0& 1
\end{array} \right)$, so that $G_I$ and $S$ generate all symplectic matrices.

\subsubsection{Decomposition of Hadamard gate into the intrinsic gate and the phase gate}
\label{app:decomposition-Hadamard-gate}

In the following, we assume that the intrinsic gate $G_I$ of Eq. \eqref{eq:intrinsicgate-symplectic} maps $Z \mapsto Z^a X^b$ with $b$ being invertible.

First, we find a Clifford gate that leaves $Z$ invariant but changes the powers of $X$ in analogy to the phase gate $S$, which allows us to control the powers of $Z$ without affecting $X$ due to
\begin{equation}
    S^l Z^a X^b S^{-l} \propto Z^{lb+a} X^b.
\end{equation}
If $b$ is invertible, we can obtain all $Z^k X^b$ with $k \in \mathbb{Z}_d$ via $l \coloneq b^{-1}(-a+k)$.

Since we would like to define a gate similar to $H S H^{-1}$, our first approach is taking $G_I S^{l_1} (G_I)^{-1}$ with $l_1$ to be determined. This gate maps $Z$ according to Eq. \eqref{eq:intrinsicgate-symplectic} via
\begin{equation}
    \begin{aligned}
        & Z \xmapsto{G_I^{-1}} Z^d X^{-b} \xmapsto{S^{l_1}} Z^{d -bl_1} X^{-b}
         \\ & \xmapsto{G_I} Z^{a(d-l_1 b)-bc} X^{ (d-l_1 b)b-bd}
        \\ & = Z^{bc+1 - a l_1 b -bc} X^{-l_1 b^2}
        =  Z^{1 - l_1 ab } X^{-l_1 b^2},
    \end{aligned}
\end{equation}
where we use that $ad = bc + 1$. Selecting $l_1 = -b^{-2} k $, we can obtain any desired power $k$ of $X$.
This, however, leaves us with $Z^{1 + k a b^{-1} }$ in addition to $X^k$.

Using afterwards the phase gate $S^{l_2}$, conjugation yields $Z^{1+k ab^{-1} + k l_2} X^k = Z^{1+k( ab^{-1} + l_2)} X^k$ which we can adjust to $Z X^k$ with the choice $l_2 = - ab^{-1}$. The total operation
\begin{equation}
    C_k \coloneq S^{-ab^{-1}} G_I S^{-b^{-2} k} G_I^{-1} 
\end{equation}
then maps $Z \mapsto Z X^k$ for any $k$. Notably, selecting $k =1$ and applying $S^{-1}$ allows us to obtain the $X$ basis.

However, the Pauli $X$ is not yet left invariant by this Clifford $C_k$ since
\begin{equation}
    \begin{aligned}
    & X \xmapsto{G_I^{-1}} Z^{-c} X^a \xmapsto{S^{-b^{-2} k}} Z^{-c - k b^{-2} a} X^a
    \\ & \xmapsto{G_I} Z^{(-c - k b^{-2} a)a + ca} X^{b(-c - k b^{-2} a) + d a}
    \\ & = Z^{- k b^{-2} a^2 } X^{1 - k b^{-1} a}
    \\ & \xmapsto{S^{-ab^{-1}}} Z^{- k b^{-2} a^2 - ab^{-1}(1 - k b^{-1} a) } X^{1 - k b^{-1} a}
    \\ & = Z^{- ab^{-1} } X^{1 - k b^{-1} a}.
    \end{aligned}
\end{equation}

Precomposing the transformation $C_k$ with $S^{l_3}$ preserves the $Z$ conjugation $Z \mapsto Z X^k$ while modifying the $X$ conjugation. Observing that $ X \mapsto X Z^{l_3}$ is transformed according to
\begin{equation}
    X \xmapsto{S^{l_3}} X Z^{l_3} \xmapsto{C_k} Z^{- ab^{-1} + l_3 } X^{1 - k b^{-1} a + kl_3},
    \label{eq:X-conjugation-C}
\end{equation}
we select $l_3 =b^{-1} a$ and in total obtain the desired transformation acting as $Z \mapsto Z X^k$ and $X \mapsto X$ via
\begin{equation}
    C_k S^{b^{-1}a} = S^{-ab^{-1}} G_I S^{-b^{-2} k} G_I^{-1} S^{a b^{-1}}.
\end{equation}

Furthermore, the choices $k=-1$ and $l_3 =  ab^{-1} + 1$ in Eq. \eqref{eq:X-conjugation-C} lead instead to $X \mapsto Z$. Followed by $S$ afterwards (which does not disrupt the $X \mapsto Z$ conjugation) results in $Z \mapsto X^{-1}$ as for the Hadamard gate. The total operation associated with the Hadamard symplectic matrix then is $S C_{-1} S^{ab^{-1} + 1}$, so
\begin{equation}
    S^{-ab^{-1}+1} G_I S^{b^{-2}} G_I^{-1} S^{a b^{-1}+1}.
    \label{eq:Hadamard-like-symplectic}
\end{equation}
This can also be verified by explicit multiplication of the symplectic matrices associated with the decomposition in Eq. \eqref{eq:Hadamard-like-symplectic}.

\subsubsection{Order of required number of measurements}

To realize an arbitrary single-qudit unitary, Eq. \eqref{eq:unitary-decomposition}, we need to implement $d^2$ generators of the type \cite{Zhou_2003, Asadian_2016}
\begin{equation}
    e^{i \sum_k \alpha_k \ket{k_P} \bra{k_P}} = A e^{i \sum_k \alpha_k \ket{k_Z} \bra{k_Z}} A^\dagger,
\end{equation}
where $A$ is the Clifford gate that maps the $Z$ basis $\{ \ket{k_Z} \}_k $ to the required Pauli eigenbasis $\{ \ket{k_P} \}_k$.

To obtain the eigenbasis $\{ \ket{k_P} \}_k$ of any $P = Z^{k_1} X^{k_2}$ with $\gcd(k_1,k_2)=1$, we need to implement the Clifford transformation associated with the symplectic matrix of $A$ in Eq. \eqref{eq:symplectic-any-pauli-basis}. Any Clifford gate requires a number of Hadamard or phase gates sub-quadratic in the dimension $d$ (for prime dimensions, even linear in $d$) \cite{Farinholt_2014}, so also $A$ and $A^\dagger$. In turn, according to Eq. \eqref{eq:Hadamard-like-symplectic}, each Hadamard symplectic matrix requires order $o_{G_I}$ (or $o_{G_I}^P$, respectively) measurements since we need to implement $G_I^{-1}$.

Thus, the number of required measurements to implement an arbitrary single-qudit unitary is of the order $d^4 o_{G_I}^P$ (more precisely, sub-quartic in $d$).
Furthermore, the Pauli order $o_{G_I}^P$ of $G_I$ necessary to obtain $G_I^{-1} = G_I^{o_{G_I} - 1}$ via measurements is at worst $d^2$ since there are only $d^2$ distinct Pauli operators $Z^{k_1} X^{k_2}$ that a Clifford operator can permute.

\subsubsection{Self-inverse intrinsic gates}
\label{app:self-inverse-intrinsic}

We now try to construct a Clifford intrinsic gate $G_I$, described via the symplectic transformation in Eq. \eqref{eq:intrinsicgate-symplectic}, that allows for universal measurement-based quantum computing, so with the element $b$ being invertible. Squaring the associated symplectic matrix, we obtain
\begin{equation}
   \left( \begin{array}{cc}
         a & c  \\
         b & d
    \end{array} \right)
    \left( \begin{array}{cc}
         a & c  \\
         b & d
    \end{array} \right)
    = \left( \begin{array}{cc}
         a^2 + cb & c(a+d)  \\
          b(a+d) & bc + d^2
    \end{array} \right) \overset{!}{=}
    \left( \begin{array}{cc} 
         1 & 0  \\
         0 & 1
    \end{array} \right).
\end{equation}
This is solved by $d = - a$ and $c = b^{-1}(1 - a^2)$. However, computing the determinant yields
\begin{equation}
    ad - bc = -a^2 - b b^{-1}(1 - a^2) = -1,
\end{equation}
so that the matrix is not symplectic unless $p=2$. For invertible $b$, the determinant condition enforces that $c = b^{-1}(ad-1)$, so that one could optimize three parameters to produce minimal Pauli order, as discussed in Sec. \ref{subsec:adapting-intrinsic-gate}.

With adjusted measurement bases on a qudit graph state of equal edge weight $w$, one obtains modified intrinsic gates $D_2 H M(w) D_1$ with $D_1$ and $D_2$ being diagonal normalizer gates most generally, so that the total operation leaves the Pauli group invariant. The transformations of $X(x)$ is then described via
\begin{equation}
    X(x) \xmapsto{D_1} X(x) Z(z_1(x)), \hspace{0.3cm} X(x) \xmapsto{D_2} X(x) Z(z_2(x))
\end{equation}
with non-linear functions $z_1, z_2: \mathbb{F}_d \mapsto \mathbb{F}_d$.
The square of the intrinsic gate $D_2 H M(w) D_1$ transforms $Z(z)$ according to
\begin{equation}
    Z(-z + z_2 (- w^{-1} (z_2 \circ z_1)(- w^{-1} z) )) X(- w^{-1} (z_2 \circ z_1)(w^{-1}z)).
\end{equation}
Thus, for $D_2 H M(w) D_1$ to be self-inverse, we require that the argument of $X$ becomes zero, so that $(z_2 \circ z_1)$ must map every argument to zero. The argument of $Z$ is then given by $-z + z_2(0)$ which cannot equal $z$ for all $z$ unless $p=2$.

In non-trivial prime-power dimensions, one might be able to find more general self-inverse normalizing gates outside of the Clifford group (so transforming the Pauli non-linearly) which optimize measurement-based quantum computing due to minimal Pauli order and which cannot be mimicked by an equal-weight qudit graph state. The full unitary normalizer is obtained by considering the isomorphism of a $p^m$-dimensional qudit and $m$ $p$-dimensional qudits \cite{heinrich2021stabiliser}, such that symplectic matrices $\left( \begin{array}{cc}
         A & C  \\
         B & D
    \end{array} \right)$,
where $A, B, C, D$ are $m \cross m$ matrices, describe the transformations of Pauli arguments $(\Vec{z}, \Vec{x})^T =(z_1, \hdots, z_m, x_1, \hdots, x_m)^T$ with entries from $\mathbb{Z}_p$. If the above maps originate from a symplectic transformation on $\mathbb{F}_{p^m}$, the matrices $A, B, C, D \in \mathbb{Z}_p^{m \cross m}$ describe the multiplication maps with $a, b, c$, $d \in \mathbb{F}_{p^m}$. However, we are interested in transformations that specifically do not originate in this way.

The finite-field Pauli gates then translate according to  \cite{heinrich2021stabiliser}
\begin{equation}
\begin{aligned}
    & Z(z) \ket{u} = \omega^{\tr(zu)} \mapsto Z(\Vec{z}) \ket{\Vec{u}} = \omega^{\Vec{z} \Vec{u}} \ket{\Vec{u}}
    \\ & X(x) \ket{u} = \ket{x+u} \mapsto X(\Vec{x}) \ket{\Vec{u}} = \ket{\Vec{x}+ \Vec{u}},
\end{aligned}
\end{equation}
so that effectively the Pauli gates factor into a tensor product of $m$ Pauli gates on $p$-dimensional systems.
Furthermore, to preserve the commutation relations on each qudit, we have the symplectic condition
\begin{equation}
\begin{aligned}
    & \left( \begin{array}{cc}
         A & C  \\
         B & D
    \end{array} \right)^T
    \left( \begin{array}{cc}
         0 & I  \\
         - I & 0
    \end{array} \right)
    \left( \begin{array}{cc}
         A & C  \\
         B & D
    \end{array} \right)
    \\ & = \left( \begin{array}{cc}
         A^T B - B^T A & A^T D - B^T C  \\
         C^T B - D^T A & C^T D - D^T C
    \end{array} \right)
    = \left( \begin{array}{cc}
         0 & I  \\
         -I & 0
    \end{array} \right).
    \label{eq:symplectic-condition}
\end{aligned}
\end{equation}
The inverse of the symplectic matrix is then given by
\begin{equation}
    \left( \begin{array}{cc}
         D^T & -C^T  \\
         -B^T & A^T
    \end{array} \right),
\end{equation}
so we see that for a self-inverse transformation, we need $A = D^T$, $B=-B^T$, and $C=-C^T$ (showing also immediately that no solution exists for $m=1$ unless $p=2$).

Hence, we effectively only need to solve for $A, B, C$,
\begin{equation}
\begin{aligned}
    & B^T + B = 0 = C^T + C, \quad A^2 + C B = I,
    \\ & A^T B + B A = 0 = C A^T + A C =0.
    \label{eq:all-equations}
\end{aligned}
\end{equation}

Furthermore, we require that $B$ is invertible, so that no $Z(z)$ gets mapped onto $X(0)$ for non-zero $z$. However, an invertible skew-symmetric matrix $B$ can only exist for $m$ even since for $m$ odd, the determinant is zero due to
\begin{equation}
    \det(B) = \det(B^T) = \det(-B) = (-1)^m \det(B).
\end{equation}
For $m$ even, an invertible skew-symmetric matrix does exist, namely, one can take the matrix with block-diagonal $\left( \begin{array}{cc}
         0 & -1   \\
          1 & 0
\end{array} \right)$ and zeroes elsewhere.
This matrix has full rank, so it is invertible and skew-symmetric by construction. The easiest solution for the whole system of equations in Eq. \eqref{eq:all-equations} is then to take $A = 0 = D$ and $C = B^{-1}$.

As an example, for $m=2$, we could take
\begin{equation}
    B =  \left( \begin{array}{cc}
          0 & -1   \\
          1 & 0
    \end{array} \right), \quad
    C =  \left( \begin{array}{cc}
         0 & 1   \\
         -1 & 0
    \end{array} \right) = B^{-1}.
\end{equation}
This transforms
\begin{equation}
\begin{aligned}
    & (z_1, z_2,x_1,x_2)^T \mapsto (x_2,-x_1,- z_2, z_1)^T.
\end{aligned}
\end{equation}
To understand such an intrinsic gate in terms of a unitary matrix, we observe that performing a swap gate, given by $((H \otimes H)CZ)^3$, followed by $H \otimes H^{-1}$, corresponds to this transformation. Hence, there exists a normalizing intrinsic gate in dimensions with $m=2$, which is self-inverse, so optimal. For $m$ even, this intrinsic gate generalizes to
\begin{equation}
    \left((H \otimes H^{-1}) ((H \otimes H)CZ)^3 \right)^{\otimes \frac{m}{2}}.
\end{equation}

Furthermore, we can map these self-inverse intrinsic gates to a diagonal entangling interaction whenever $B$ is invertible, since then $Z(\Vec{z})$ gets mapped onto $X(B \Vec{z})$ with non-zero $B \Vec{z}$ for all $\Vec{z}$. This means that $G_I \ket{k_Z}$ is an eigenstate of a Pauli with non-zero $X$, so an equal superposition state, such that all matrix entries of $G_I$, $ \bra{j_Z} G_I \ket{k_Z}$, have equal modulus. Thus, the phases can be written into a diagonal entangling gate $G_E$, Eq. \eqref{eq:diagonal-entangling}, and resource qudits initialized in $\ket{+}$, as discussed in Sec. \ref{sec:diagonal-entangling-gates}.

\subsection{Resource state geometry for two-qudit entangling gate realization}
\label{app:resource-graph-manipulations}

So far, we have focused on one-dimensional resource states, Fig. \ref{fig:setting} $(b)$, that can support the implementation of any single-qudit unitary. For universal measurement-based quantum computing, we, however, additionally, require the ability to perform a two-qudit entangling gate, Fig. \ref{fig:setting} $(c)$, so a universal two-dimensional resource structure that can be tailored to a desired quantum circuit via single-qudit measurements, Fig. \ref{fig:setting} $(d)$.

For graph states, manipulation rules via Pauli measurements are known \cite{hein2006entanglementgraphstatesapplications}, which are convenient to cut out the elementary building blocks of Figs. \ref{fig:setting} $(b)$ and $(c)$. This strategy turns out to work also for resources prepared via diagonal Clifford entangling gates discussed first since these resource states are locally Clifford equivalent to graph states. Instead, for block-diagonal entangling gates, we use a less symmetric resource state geometry that relies on mediator qudits to either disconnect or entangle two horizontal computational lines.

\subsubsection{Diagonal Clifford entangling gates}
\label{app:LC-equivalent-cluster-state}

We assume resources, characterized by diagonal Clifford entangling gates $G_E = \sum_{jk} e^{i \theta_{jk}} \ket{jk} \bra{jk}$. Such resource states can be visualized with directed graphs, where vertices correspond to qudits in $\ket{+}$ and an edge corresponds to $G_E$ being applied. This case includes the standard controlled-phase gate and the ionic light-shift gate \cite{nativequditentanglinggate}.

Since the controlled-phase gate $CZ$ is the only diagonal two-qudit Clifford group generator, we find that any Clifford diagonal entangling gate can be expressed via
\begin{equation}
\begin{aligned}
    & (C_1 \otimes C_2) CZ^N
\end{aligned}
\end{equation}
where $C_1$ and $C_2$ are single-qudit diagonal Clifford gates and $N \in \mathbb{Z}_d$ is a non-zero integer. For finite-field qudits, we have $(C_1 \otimes C_2) M(N) CZ M(N)^\dagger$ with $N \in \mathbb{F}_d \backslash \{0 \}$ instead. If the entangling gate is part of the unitary normaliser (so it preserves the Pauli group) but not the finite-field Clifford group, we replace every qudit with lower-dimensional systems and consider the corresponding $CZ$ gates between qudit pairs together with local diagonal Clifford gates.
Hence, the resulting resource state is locally Clifford equivalent to a graph state with the same entanglement structure, and if $N=1$, it is locally Clifford equivalent to the cluster state. We have seen this kind of local Clifford equivalence of the entangling gate to $CZ$ for the light-shift gate resources in Sec. \ref{sec:lightshift-resources}.
The resource state can then be manipulated in the same way as graph states, taking the local Clifford operations $C_1$ and $C_2$ into account, in particular, vertices can be deleted and edges created, so that one can use two-dimensional lattices as a universal resource states, as shown in Fig. \ref{fig:2D-resource-diagonal-gates}.

Observing how $G_E$ conjugates $X \otimes I$ and $I \otimes X$, the resource states can be visualized with directed graphs to specify the direction of $G_E$ being applied. The stabilizer group generators associated with each vertex $v$ then become
\begin{equation}
\begin{aligned}
    & P_v(|N_v^i|,|N_v^o|) \prod_{u \in N_v} Z_u^{N},
\end{aligned}
    \label{eq:stabilizer-diagonal-gate}
\end{equation}
where $N_v^i$ is the neighborhood of vertex $v$ with incoming edges, $N_v^o$ the one with outgoing edges, and
\begin{equation}
    P_v(|N_v^i|,|N_v^o|) = C_1^{|N_v^o|} C_2^{|N_v^i|} X_v \left(C_1^{|N_v^o|} C_2^{|N_v^i|} \right)^\dagger,
\end{equation}
so $X_v$ multiplied with a $Z_v$ gate to some power and a phase, potentially.
For prime-power-dimensional qudits, we instead have the generators
\begin{equation}
\begin{aligned}
    & P_v(|N_v^i|,|N_v^o|,x) \prod_{u \in N_v} Z_v(Nx)
\end{aligned}
\end{equation}
for all $x \in \mathbb{F}_d$, associated with each vertex $v$.

Even without specifying the resource state stabilizer group generators explicitly, we can already outline a procedure for vertex removal and removal with simultaneous edge creation.
To remove a vertex $v$, we can measure it in the $Z_v$ basis since the measurement outcome will commute with all neighbor stabilizers but not with the vertex stabilizer, which will be replaced with the stabilizer specifying the outcome of the measurement. Since the vertex is removed together with the edges after the measurement, one needs to apply an operation onto the neighbor stabilizers that adjusts $|N^o|$ or $|N^i|$ to one neighbor less without modifying $Z$. This operation is $C_1^\dagger$ and $C_2^\dagger$, respectively.

For the local complementation measurement, we can measure a vertex in the basis $P_v(|N_v^i|,|N_v^o|) Z_v^N$ (or $P_v(|N_v^i|,|N_v^o|,x) Z_v(Nx)$, respectively). This measurement does not commute with any of the involved vertex stabilizers, however, one can choose the measured vertex stabilizer as the stabilizer to be replaced with the measurement outcome and multiply the neighbor stabilizers with it to keep them as commuting stabilizers. Hence, any neighbor vertex gains the neighbors of the measured vertex and itself as neighbors. One then needs to remove the self-edge by applying $S^{-N}$. Furthermore, to preserve the structure of Eq. \eqref{eq:stabilizer-diagonal-gate}, one needs to adjust the Pauli $P_u(|N_u^i|,|N_u^o|)$ for each neighbor $u$ of vertex $v$ to the correct count of incoming and outgoing edges. In this way, one can also select the directions of the new edges. Undesired phases can then be removed by applying appropriate Pauli operators.

\subsubsection{Block-diagonal Clifford entangling gates}
\label{app:2Dresource-blockdiagonal}

According to the Clifford group generators, any two-qudit Clifford gate can be expressed as a product of local Clifford gates and $CZ$ gates. Since we restrict ourselves to block-diagonal gates, only diagonal single-qudit Clifford gates on the control qudit preserve the block-diagonal structure, whereas the local Clifford acting on the target can be arbitrary. Hence, in general, we can express a block-diagonal Clifford entangling gate via
\begin{equation}
    (C_1 \otimes C_2) CP,
\end{equation}
where $CP$ is a controlled-Pauli, $C_1$ a diagonal local Clifford gate, and $C_2$ an arbitrary single-qudit Clifford gate.

Given a controlled-Pauli entangling gate $CP$, the prior transport of the quantum information with the two-dimensional resource state in Fig. \ref{fig:2D-resource-blockdiagonal} no longer commutes with the vertical lines as it has been the case for diagonal gates; however, the posterior transport commutes. Hence, one can consider the simplified scenario of Fig. \ref{fig:mediator-qudit} of having transported the two-qudit quantum state, which one either wants to transport further or entangle via the mediator qudit.

For the Pauli $P$, we can always find a Clifford $C$ that maps $P$ to $Z^l$ for some power $l$ up to a phase $e^{i \eta }$. This follows from the fact that any Pauli $Z^m X^n$ can be mapped to $Z^{\gcd(m,n)}$, where $\gcd(m,n) \eqcolon l$ is the greatest common divisor of $m$ and $n$ \cite{Farinholt_2014}.
This means that any controlled-Pauli with $P \propto Z^a X^b$ can be expressed via
\begin{equation}
    (I \otimes C^{-1}) CZ^{l} (I \otimes C).
    \label{eq:controlled-Paulis}
\end{equation}

Then we can initialize a mediator qudit in $\ket{\phi} = C^{-1} \ket{0_X}$, so that we obtain
\begin{equation}
    \begin{aligned}
        & CP \left( \sum_{k} \alpha_k \ket{k} \ket{\phi} \right) = \sum_{k} \alpha_k \ket{k} C^{-1} e^{i \eta k} Z^{lk} \ket{0_X}
        \\ & = \sum_{k} \alpha_k e^{i \eta k} \ket{k} C^{-1} H \ket{(lk)_Z} = \sum_{k} \alpha_k D_{\eta} \ket{k} G_C \ket{k_Z},
    \end{aligned}
\end{equation}
where we define
\begin{equation}
    G_C \coloneq C^{-1} H M(l), \quad D_{\eta} \coloneq \sum_k e^{i \eta k} \ket{k} \bra{k}.
\end{equation}
Notably, the gate $G_C$ only defines a unitary operation for $M(l)$ with $l$ being invertible (which is always true for non-zero $l$ in the case of finite-field qudits). For instance, for the $CZ$ gate, we have $C=I_d$ and $l=1$, so $M(1) = I_d$, and we arrive at standard cluster state quantum computing. In turn, for $CX$ this initialization leads to $C=H$ and $l=1$, so that $G_C = I_d$ and $\ket{\phi} = H^{-1} \ket{0_X} = \ket{0_Z}$. If we only want to mediate entanglement and not perform single-qudit gates, the initialization $\ket{0_Z}$ for the $CX$ gate suffices. However, for single-qudit unitary implementation, one would require an initialization with a non-trivial intrinsic gate $G_C$. Therefore, we choose a different initialization approach in Sec. \ref{subsec:CX-gate-preparation} for the computational qudits.

Applying a controlled-Pauli entangling gate twice with two different controls and the mediator qudit in $\ket{\phi} = C^{-1} \ket{0_X}$ as the common target then results in
\begin{equation}
    \sum_{k,j} D_{\eta} \ket{k} \bra{k} \otimes G_C \ket{j+k} \otimes D_{\eta} \ket{j} \bra{j}.
\end{equation}
We can then measure the mediator qudit in the basis $ \{ G_C \ket{k_X} \}_k =  \{ G_C Z^k \ket{0_X} \}_k$ to disconnect the computational qudits of $\ket{\psi}$ or in the basis $ \{ G_C S^{-1} \ket{k_X} \}_k$ if we want to instead entangle both qudits via $CZ (S \otimes S )$ (up to a by-product of $Z^{-k} \otimes Z^{-k}$ in both cases). The gates $D_{\eta}$ on each computational qudit can be removed in the next computational step by adjusting the measurement basis appropriately.

So far, we understand how to construct the resource states and choose measurement bases on an intermediate mediator qudit for controlled-Pauli $CP$ entangling gates using the resource state depicted in Fig. \ref{fig:2D-resource-blockdiagonal}.
A diagonal Clifford gate $C_1$ in Eq. \eqref{eq:structure-blockdiagonal} does not affect the general strategy pursued for the $CP$ entangling gate. However, a non-trivial $C_2$ gate that does not commute with $CP$ generally disturbs the property $P^k P^j = P^{k+j}$, which we have implicitly used to pick appropriate measurement bases that entangle or disentangle the two-qudit input.

For resources that are prepared via the application of the respective entangling gates, we can remedy this by removing the $C_2$ gate on the target qudit after the application of the first block-diagonal entangling Clifford gate, so by applying $C_2^{-1}$.

\subsection{Local Clifford equivalence of stabilizer state resources and qudit graph states}
\label{app:CX-weight-1-graph}

All stabilizer states of prime-power dimension are locally Clifford equivalent to a qudit graph state. If the equivalence is given by diagonal single-qudit Cliffords and the entanglement connectivity is preserved, a qudit graph state can mimic the stabilizer state resource by adjusting the measurement bases for two subsequent measurements.

After initialization in $\ket{\varphi}$ and applying $C$ from Eq. \eqref{eq:controlled-Paulis}, the resource qudits are in an Pauli eigenstate. In turn, for finite-field qudits of prime-power dimension, all Pauli eigenstates are contained within the set $\{ Z(1), XZ(a) \mid a \in \mathbb{F}_d \}$. If the resource qudit becomes a $Z$ eigenstate, no entangling interaction is performed; this case should be avoided, which is why we introduced a non-$\ket{+}$ initialization for $CX$-gate resources.
If the resource qudit is in one of the other eigenstates, we can express it via $S(k) \ket{+}$, so that the intrinsic gate becomes $C^{-1} S(k) H M(l)$. Whenever this intrinsic gate can be expressed via $S(s_2) H M(w) S(s_1)$, a qudit graph state can imitate the resulting stabilizer state resource via adjusting measurement bases, as discussed in Sec. \ref{subsec:adapting-intrinsic-gate}. For instance, for $CX$-gate resources, we have $l=1$, $C^{-1} = H^{-1}$ and $\ket{\varphi} = S(1) \ket{0_X}$, so that the resource qudit becomes $H S(1) \ket{0_X}$, which can be re-expressed via $S(-1) \ket{0_X}$. Hence, the intrinsic gate becomes $H^{-1} S(-1) H$, which transforms the Pauli operators in the same way as $S(1) H^{-1} S(1) = S(1) H M(-1) S(1)$, resulting in a qudit graph state one-dimensional chain with edge weight $w = -1$ being able to imitate it. In fact, even the two-dimensional structure of the $CX$-gate resource with mediator qudits is locally Clifford equivalent to a qudit graph state of edge weight $-1$, as argued in the following.

We first consider a one-dimensional resource state chain and the associated stabilizer group generators. Initially, each resource qudit $v$ is in $\ket{+}$ with stabilizer $X$, onto which an $S$ gate and sequentially $CX$ gates are applied, resulting in the generator
\begin{equation}
    g_v = \tau X_v Z_v Z_{v-1}^{-1} \prod_{u \in \{v+1, \hdots ,n \} } X_u
\end{equation}
for each vertex $v$ (we consider the integer-ring formalism for the calculation), where we label the vertices with integers in $\{ 1, \hdots, n\}$ for a chain of length $n$. Here, we used that $CX $ conjugates $I \otimes Z \xmapsto{CX} Z^{-1} \otimes Z$ and $X \otimes I \xmapsto{CX} X \otimes X$ while $Z \otimes I$ and $I \otimes X$ are left invariant. To localize these stabilizer group generators, as for graph state stabilizers, we multiply $g_v$ with $g_{v+1}^{-1}$ for each $v < n$, so that the support from qubits in $\{v+2, \hdots, n\}$ is canceled, and we obtain the alternative stabilizer group generators
\begin{equation}
\begin{aligned}
    & g_v' = g_v g_{v+1}^{-1} = X_v Z_v Z_{v-1}^{-1} X_{v+1} Z_{v} Z_{v+1}^{-1} X_{v+1}^{-1}
    \\ & = X_v Z_v^2 Z_{v-1}^{-1} X_{v+1} Z_{v+1}^{-1} X_{v+1}^{-1} 
    = \omega X_v Z_v^2 Z_{v-1}^{-1} Z_{v+1}^{-1}.
    \label{eq:new-stabilizers-linearCX}
\end{aligned}
\end{equation}
For the last qudit in the chain, $g_n' = g_n$. Thus, up to an $S^2$ gate (which equals $Z$ for $p=2$) on all qudits with $v < n$ and an $S$ gate on the last qudit, one obtains the vertex stabilizers $Z_{v-1}^{-1} X_v Z_{v+1}^{-1}$, a linear graph state with edge weight $w =-1$.

Considering a two-dimensional resource state structure with mediator qudits  $m$, where the mediator qudits are the targets of both $CX$ gates, yields the mediator stabilizer group generators
\begin{equation}
    g_m = \tau X_m Z_m Z_{m-1}^{-1} Z_{m+1}^{-1},
\end{equation}
whereas the neighbors $m$ and $m-1$ of the mediator have the stabilizers $g_v'$ of Eq. \eqref{eq:new-stabilizers-linearCX} with one additional $CX$ gate between the neighbor and mediator, so
\begin{equation}
    \omega X_v Z_v^2 Z_{v-1}^{-1} Z_{v+1}^{-1} X_m.
\end{equation}
Here, $v-1$ and $v+1$ are the neighbors of the mediator neighbor $v$.
Applying $H^{-1}$ on the mediator, we obtain
\begin{equation}
    g_m' = \tau Z_m^{-1} X_m Z_{m-1}^{-1} Z_{m+1}^{-1},
\end{equation}
followed by an $S$ gate on the mediator,
\begin{equation}
    g_m'' = \omega Z_m^{-1} X_m Z_m Z_{m-1}^{-1} Z_{m+1}^{-1} = X_m Z_{m-1}^{-1} Z_{m+1}^{-1},
\end{equation}
yielding a graph state stabilizer on the mediator. On the neighbors of the mediator, after an $H^{-1}$ gate and a subsequent $S$ gate on the mediator, their stabilizer group generators become
\begin{equation}
    \omega X_v Z_v^2 Z_{v-1}^{-1} Z_{v+1}^{-1} Z_m^{-1}.
\end{equation}
Hence, up to $S^2$ on the one-dimensional resource state chain ($S$ for the last qudit of each chain) and $S H^{-1}$ on the mediators, we recover a qudit graph state with edge weight $w=-1$ also for the two-dimensional structure.

\subsection{Clifford circuits via simultaneous measurements}
\label{app:clifford-circuits}

For cluster state qudits, which are associated with the Hadamard gate as an intrinsic operation, Clifford circuits can be executed with simultaneous measurements without the need to adapt measurement bases. This follows from the Hadamard gate and the phase gate generating the single-qudit Clifford group \cite{Farinholt_2014}.
The multiplication gate, which is often mentioned as a Clifford group generator both for finite-field \cite{heinrich2021stabiliser} and integer-ring qudits \cite{LinearizedStabilizerFormalism}, is redundant in a minimal generating set since it can be expressed via the Hadamard and phase gates \cite{Farinholt_2014}. Namely, the symplectic matrix associated with the multiplication gate $M(\lambda) \mapsto \left( \begin{array}{cc}
    \lambda^{-1} & 0  \\
     0 & \lambda
\end{array} \right)$, can be decomposed according to \cite{Farinholt_2014}
\begin{equation}
    H S(\lambda) H S(\lambda^{-1}) H S(\lambda).
\end{equation}
For integer-ring qudits, $S(\lambda)$ becomes $S_d^\lambda$, whereas for finite-field qudits with odd dimension, it is $S_d^F(\lambda)$. To obtain the same symplectic matrix as of $S_d^F(\lambda)$ in even prime-power dimensions, we can replace $S(\lambda)$ with $M(l^{-1}) S_d^F M(l)$, where $l^2 = \lambda$ (such an $l$ always exists since the map $l \mapsto l^2$ is a bijection in $\mathbb{F}_{2^m}$). This decomposition is given in Ref. \cite{Farinholt_2014} for integer-ring qudits with the convention that the symplectic vector $(a,b)$ corresponds to the Pauli $X_d^a Z_d^b$ instead of our convention $Z_d^a X_d^b$.

For our generalized qudit stabilizer state resources, we can only claim that Clifford circuits can be realized with simultaneous measurements without the need to adapt measurement bases if we can decompose all Clifford group generators into a measurement pattern of intrinsic gates $G_I$ and diagonal Clifford gates, allowing us to propagate Pauli by-products for any measurement outcome.
Diagonal gates such as the phase gate $S(\lambda)$ are easily implemented in measurement-based quantum computing, and we have shown in Appendix \ref{app:decomposition-Hadamard-gate} that the Hadamard gate can be expressed via a combination of the intrinsic gate $G_I$ and phase gates. Therefore, every single-qudit Clifford gate can be decomposed into Clifford $G_I$ and phase gates, which can then be implemented in a single time step without the need to adjust measurement bases. Since the two-qudit entangling gate that can be implemented on a two-dimensional resource state is also Clifford, we can then realize the whole Clifford group.

Note that neither the decomposition of a Hadamard gate nor the multiplication gate is necessarily optimal, and a shorter Clifford measurement pattern may be available.

\end{appendix}

\end{document}